\documentclass[
 rreprint,
 amsmath,amssymb,
 aps,
 superscriptaddress
]{revtex4-2}

\usepackage{graphicx}
\usepackage{tikz}
\usepackage{epstopdf}
\usepackage{float}
\usepackage{placeins}
\usepackage{soul}
\sethlcolor{yellow}
\usepackage{dcolumn}
\usepackage{bm}
\usepackage{subcaption} 
\usepackage{xcolor}

\begin{document}

\preprint{APS/123-QED}

\title{A Unified Description of Dirac-Cone Anisotropies in Two Dimensions}%

\author{L. C. T. Brito}
\email{lcbrito@ufla.br}
\affiliation{Departamento de Física, Instituto de Ciências Naturais,
Universidade Federal de Lavras, Caixa Postal 3037,
37200-900, Lavras, Minas Gerais, Brazil}

\author{Cleverson Filgueiras}
\email{cleverson.filgueiras@ufla.br}
\affiliation{Departamento de Física, Instituto de Ciências Naturais,
Universidade Federal de Lavras, Caixa Postal 3037,
37200-900, Lavras, Minas Gerais, Brazil}

\author{D. M. Lopes}
\email{daniel.lopes2@estudante.ufla.br}
\affiliation{Departamento de Física, Instituto de Ciências Naturais,
Universidade Federal de Lavras, Caixa Postal 3037,
37200-900, Lavras, Minas Gerais, Brazil}

\author{A. G. Martins}
\email{andrey\_martins@uepa.br}
\affiliation{Departamento de Física, Centro de Ciências Sociais e Educação, Universidade do Estado do Pará, 66050-540, Belém, Pará, Brazil}

\author{Igor S. S. de Oliveira}
\email{igor.oliveira@ufla.br}
\affiliation{Departamento de Física, Instituto de Ciências Naturais,
Universidade Federal de Lavras, Caixa Postal 3037,
37200-900, Lavras, Minas Gerais, Brazil}

\date{\today}

\begin{abstract}
Anisotropies in two-dimensional materials are responsible for a variety of effects that significantly modify their physical properties. In this paper, we present a covariant modification of the Dirac equation that incorporates anisotropies into the effective low-energy description around the Dirac point. The model is constructed by analogy with a Lorentz-violating extension of the Standard Model of elementary particles and yields a (2+1)-dimensional framework describing physical effects such as shifted, tilted, and distorted Dirac cones through its free parameters. The proposed model thus provides a general and unified framework in which distinct anisotropy-induced modifications of the Dirac spectrum and their combinations can be systematically characterized. We further apply the model to strained graphene and show that its effective parameters can be quantitatively extracted from first-principles electronic band structures while retaining a direct geometrical interpretation. These results establish a connection between the microscopic electronic structure and a material-independent effective description of anisotropic two-dimensional Dirac systems.
\end{abstract}

\maketitle


\section{Introduction}

Research interest in two-dimensional (2D) materials has grown rapidly and steadily over the last two decades, mainly because of their potential technological applications~\cite{AJAYAN,BUTLER}. From a scientific perspective, the development of experimental techniques capable of producing ultrathin materials in controlled environments has made it possible to observe planar effects characteristic of truly two-dimensional systems~\cite{NOVOSELOV3}. As a result, the behavior of electrons and quasiparticles in these materials can be effectively described as if they were confined to a two-dimensional world. Notable examples include graphene, silicene~\cite{VOON}, and transition metal dichalcogenides such as MoS\textsubscript{2}~\cite{JIANG}, which exhibit a variety of electronic behaviors under suitable conditions.

From a theoretical standpoint, interest in planar physics is not new. For example, the possibility of exotic statistics in two dimensions was reported in the classical 1977 paper by Leinaas and Myrheim~\cite{LEINAAS}. This idea was later explored in quantum field theory, leading to the concept of anyons~\cite{WILCZEK}—quasiparticles whose statistics are governed not by the usual permutation symmetry of bosons and fermions, but by the braid group~\cite{WU}. These developments highlight the richness of two-dimensional systems, not only from a technological perspective but also as platforms for investigating fundamental aspects of quantum theory. The existence of anyons and their nontrivial statistics has recently been confirmed experimentally~\cite{NAKAMURA,BARTOLOMEI}.

Another development, now more directly related to the present work, is the seminal paper by Semenoff, one of the first to propose that a Dirac-like equation could emerge as an effective description of electrons in a honeycomb lattice~\cite{SEMENOFF}. In the low-energy regime, the system is described around two degeneracy points in the Brillouin zone—now known as Dirac points~\cite{CASTRO}—where two species of relativistic (2+1)-dimensional fermions emerge. This result is remarkable because it predates the isolation of graphene by nearly two decades~\cite{NOVOSELOV2}. It is now well established that this effective description extends beyond graphene, defining a broader class of systems known as Dirac materials~\cite{WEHLING,GOERBIG}.

In the absence of anisotropies, the low-energy excitations around the Dirac points are well described by massless Dirac fermions in (2+1) dimensions, particularly in materials such as graphene~\cite{CASTRO}. The corresponding dispersion relation forms gapless Dirac cones around the Dirac points. In this work, we focus on two-dimensional Dirac materials whose physical properties are modified by anisotropic effects. These effects manifest directly in the dispersion relations, altering the Dirac cones in different ways~\cite{WANG}. In the absence of external fields, anisotropies in these materials typically arise either intrinsically from the crystal structure or emergently from external mechanical deformations.

Strain-induced anisotropies~\cite{PEREIRA,DEJUAN,AMORIM,NAUMIS} are perhaps the most physically relevant and experimentally accessible cases for the present work. The observed effects include tilted Dirac cones~\cite{GOERBIG2,MILIĆEVIĆ,ALMARZOOG,MOJARRORAMIREZ,MOJARRO}, elliptical cones~\cite{IUROV} and band-gap opening~\cite{GUI,PEREIRA2}. These phenomena have been investigated using both effective Hamiltonian approaches and first-principles computational methods, which provide microscopic support and enable material-specific predictions.

In this contribution, we present an anisotropic extension of the Dirac equation in (2+1) dimensions, derived from a covariant Lagrangian, that encompasses key effects such as shifted, tilted, and distorted Dirac cones, as well as mass-gap generation. The proposed model provides a unified theoretical framework for describing a broad class of anisotropy-induced modifications of the dispersion relation, independent of material-specific microscopic details, thereby offering a natural setting for phenomenological investigations. Moreover, the model explicitly incorporates a defining property of Dirac materials: the relativistic-like behavior of quasiparticles in the low-energy regime near the Dirac points. Notably, our approach mirrors the general structure of effective field theories commonly employed in high-energy physics~\cite{CAMBIASO}, and which have recently been applied to Dirac and Weyl semimetals in (3+1) dimensions~\cite{KOSTELECKÝ}. Here, we propose an effective model in which the full Lagrangian formalism of quantum field theory can be implemented in (2+1) dimensions, opening new possibilities for investigating quantum phenomena in two-dimensional systems~\cite{VOZMEDIANO}.

The organization of the paper is as follows. In Sec.~\ref{DiracMaterials}, we review Dirac materials, highlighting the existence of Dirac points and the symmetries relating them. In Sec.~\ref{AnisotropicDirac}, we present the effective model for anisotropic Dirac materials and derive the corresponding dispersion relation. Section~\ref{AnisotropicEffects} analyzes this dispersion relation for specific choices of the model parameters, recovering several anisotropic effects observed in real materials, such as Dirac-cone shifts, tilting, elliptic distortions, and mass-gap generation. Section~\ref{validation} applies the model to strained graphene using first-principles calculations, providing a quantitative benchmark of the effective description and illustrating its usefulness for phenomenological investigations. Finally, Sec.~\ref{conclusion} summarizes the main results and conclusions, including a brief discussion of possible extensions and implications. Two appendices provide additional details supporting the semimetallic character of the model in the massless limit and describing the computational method employed in the application presented in Sec.~\ref{validation}. Throughout this work, we adopt the Minkowski metric $\eta_{\mu\nu}$ with signature $(1,-1,-1)$ and assume Einstein's summation convention.

\section{Dirac materials}
\label{DiracMaterials}

In Dirac materials, the low-energy behavior of quasiparticles around the Dirac points is described by the massless Dirac equation, which in covariant form reads
\begin{equation}
i\hbar v_{F}\gamma^{\mu}\partial_{\mu}\psi=0,
\label{massless}
\end{equation}
where \(v_{F}\) is the Fermi velocity and \(\gamma^{\mu} = (\gamma^{0},\gamma^{1},\gamma^{2})\) are the Dirac matrices satisfying the algebra
\begin{equation}
 \left\{ \gamma^{\mu},\gamma^{\nu}\right\}  = 2 \eta^{\mu\nu} I.
 \label{DiracAlgebra}
 \end{equation}
For the purposes of the present work, we adopt the following \((2+1)\)-dimensional representation in terms of Pauli matrices:
\begin{equation}
\gamma^{0}=\sigma_{3},\,\,\,\,\,\,\gamma^{1}=i\sigma_{1},\,\,\,\,\,\,\text{and}\,\,\,\,\,\,\gamma^{2}=i\sigma_{2},
\label{repres1}
\end{equation}
which satisfies
\begin{equation}
\left[\gamma^{\mu},\gamma^{\nu}\right]=2i\epsilon^{\mu\alpha\nu}\gamma_{\alpha}.
\end{equation}
The derivative operator is defined as usual by
\[
\partial_\mu = \left(\frac{1}{v_{F}}\frac{\partial}{\partial t}, \nabla\right).
\]

In \((2+1)\) dimensions, there exists another representation of the gamma matrices that differs from (\ref{repres1}) by the sign of one of the matrices. These two representations are inequivalent, which is a general feature of odd-dimensional space-times~\cite{SHIMIZU}. In the context of Dirac materials, this structure has an important physical interpretation, since the two representations are associated with the Dirac points \(K\) and \(K'\), i.e., the degeneracy points where the Dirac cones are located \cite{CASTRO}. In graphene, for example, these points lie at the corners of the Brillouin zone.

The Dirac equation (\ref{massless}) leads to the Dirac Hamiltonian
\begin{equation}
H=v_{F}\vec{\alpha}\cdot\vec{p},
\label{hamiltonian}
\end{equation}
where
\begin{equation}
\vec{\alpha}=\left(\alpha_{x},\alpha_{y}\right)=\left(i\sigma_{2},i\eta\sigma_{1}\right).
\end{equation}
The parameter \(\eta = \pm 1\) identifies the two inequivalent representations of the gamma matrices introduced above. Note that these representations are connected by the transformation
\begin{equation}
(p_{x},p_{y}) \rightarrow (p_{x},- p_{y}),
\end{equation}
which corresponds to a parity transformation in \((2+1)\) dimensions~\cite{SHIMIZU,JACKIW,DUNNE}. Physically, this means that the two Dirac points \(K\) and \(K'\) are related by parity. Consequently, a parity-invariant low-energy theory is obtained only when both representations are included~\cite{BASHIR}. In graphene, for example, this structure emerges naturally from the tight-binding approach~\cite{SEMENOFF,CASTRO}. More generally, in two-dimensional Dirac materials the existence of degenerate points \(K\) and \(K'\) follows from symmetries that depend on the specific material~\cite{WEHLING}, but which emerge at low energies as a parity-conserving theory of two species of Dirac fermions, each associated with a distinct representation of the gamma matrices. In what follows, we work in the specific representation \(\eta = 1\), focusing on a single Dirac point.

\section{Anisotropic Dirac equation and the dispersion relation}
\label{AnisotropicDirac}

In real 2D Dirac materials, anisotropic effects—arising from strain~\cite{AMORIM,NAUMIS}, substrate interactions~\cite{WEHLING,WANG}, or intrinsic lattice deformations~\cite{AJAYAN,WANG,KIM}—can substantially modify the physical properties of low-energy quasiparticles. These effects often manifest in the dispersion relation near the Dirac points and may lead to modifications of the Dirac cones. We propose that, in the presence of  anisotropic effects, the quasiparticles around the Dirac points can be effectively described at low energies by modifying the massless Dirac equation in Eq.~(\ref{massless}) as
\begin{equation}
i\hbar v_{F}\Gamma^{\mu}\partial_{\mu}\psi-Mv_{F}^{2}\psi=0, 
\label{modEq}
\end{equation}
where
\begin{equation}
\Gamma^{\mu}= \gamma^{\mu}+a_{\,\,\nu}^{\mu}\gamma^{\nu}+b^{\mu}I
\label{mod1}
\end{equation}
and
\begin{equation}
M= m + d_{\mu}\gamma^{\mu}.
\label{mod2}
\end{equation}
The parameters \(m\), \(a_{\,\,\nu}^{\mu}\), \(b^\mu = (b^{0},\vec{b}) = (b^{0}, b^{1},b^{2})\), and \(d^\mu = (d^0 , \vec{d}) = (d^0 , d^{1},d^{2})\) are constants. We note that, like the parameter \(m\), the parameter \(d^\mu\) has dimensions of mass, whereas \(a^{\mu}_{\,\,\nu}\) and \(b^{\mu}\) are dimensionless.  The general structure of \(\Gamma^{\mu}\) and \(M\) in Eqs.~(\ref{mod1}) and (\ref{mod2}) follows from the fact that the identity \(I\) and the Pauli matrices together form a basis for all \(2\times 2\) matrices. This \((2+1)\)-dimensional construction parallels the extensions considered in \((3+1)\) dimensions, where the modifications are expressed in terms of the \(4\times 4\) matrix basis~\cite{CAMBIASO,KOSTELECKÝ}. 

It is worth noting that Eq.~(\ref{modEq}) can be derived from the Lagrangian density
\begin{equation}
\mathcal{L}=i\hbar v_{F} \bar{\psi}\Gamma^{\mu}\partial_{\mu}\psi- v_{F}^{2}\bar{\psi} M \psi.
\label{lagrangian}
\end{equation}
 Thus, we propose a covariant generalization of the Dirac equation that allows anisotropic features of two-dimensional Dirac materials to be incorporated within a field-theoretic framework. In particular, it is well known that in \((2+1)\) dimensions the mass term in the Lagrangian (\ref{lagrangian}) breaks parity~\cite{JACKIW,DUNNE}, thereby opening a mass gap in the dispersion relation. 

As usual, the dispersion relation can be obtained from the modified equation by working in momentum space, where the momentum in \((2+1)\) dimensions is written as \(p^{\mu} = (p^{0},\vec{p}) = (p^{0},p^{1},p^{2})\), with \(p^{0} = \frac{E}{v_F}\). In this representation, Eq.~(\ref{modEq}) takes the form
\begin{equation}
\left(\Gamma^{\mu}p_{\mu}-Mv_{F}\right)\psi=0.
\label{DiracMoment}
\end{equation}
Since the extended Dirac operator \(\Gamma^{\mu}p_{\mu}-M v_{F}\) must be singular for nontrivial solutions of the Dirac equation to exist, we impose the usual condition
\begin{equation}
\det\left(\Gamma^{\mu}p_{\mu}-M v_{F}\right)= 0.
\label{detCondition}
\end{equation}
Using the definitions given in Eqs.~(\ref{mod1}) and (\ref{mod2}), the expression inside the determinant in Eq.~(\ref{detCondition}) can be written as
\begin{equation}
\Gamma^{\mu}p_{\mu}-Mv_{F} = \left(p_{\mu}-v_{F}d_{\mu}+a_{\,\,\mu}^{\nu}p_{\nu}\right)\gamma^{\mu}+\left(p^{\mu}b_{\mu}-mv_{F}\right)I.
\label{operator}
\end{equation}
Here and in what follows, we use the notation \(p^{\mu}b_{\mu} = \eta_{\mu\nu} p^{\mu} b^{\nu} = p^0 b^0 - \vec{p} \cdot \vec{b}\), where the last term denotes the standard scalar product in two-dimensional Euclidean space.

For a general \(2\times2\) matrix \(A\), the determinant can be expressed in terms of traces as
\begin{equation}
\det A=\frac{1}{2}\left[\text{Tr}^{2}\left(A\right)-\text{Tr}\left(A^{2}\right)\right].
\end{equation}
Thus, using Eq.~(\ref{operator}), Eq.~(\ref{detCondition}) becomes
\begin{equation}
p^{2}+p^{\mu}p^{\nu}\Upsilon_{\mu\nu}-2v_{F}\Theta_{\mu}p^{\mu}-v_{F}^{2}\left(m^{2}-d^{2}\right)=0,
\label{dispersionRel}
\end{equation}
where we have defined \(d^{2} = \left(d^{0}\right)^2 - \vec{d}^2\),
\begin{equation}
\Upsilon_{\mu\nu}=2a_{\mu\nu}+a_{\mu\tau}a_{\nu\sigma}\eta^{\sigma\tau}-b_{\mu}b_{\nu}
\label{upsilon}
\end{equation}
and
\begin{equation}
\Theta_{\mu}=d_{\mu}+a_{\mu\rho}d^{\rho}-mb_{\mu}.
\label{theta}
\end{equation}
In the algebra leading to Eq.~(\ref{dispersionRel}), we used the Dirac algebra given in Eq.~(\ref{DiracAlgebra}), together with
\begin{equation}
 \text{Tr}\left(\gamma^{\mu}\right) = 0\,\,\,\,\,\,\,\, \text{and} \,\,\,\,\,\,\,\,\text{Tr}\left(I\right) = 2.
\end{equation}

As can be seen from Eq.~\eqref{dispersionRel}, the vector coefficient $b^{\mu}$ enters the covariant dispersion relation only through the combinations defined in Eqs.~\eqref{upsilon} and \eqref{theta}. This indicates that this coefficient does not introduce an independent geometric deformation of the Dirac cones. Instead, its contributions modify the same effective parameters $\Upsilon_{\mu\nu}$ and $\Theta_\mu$ that already determine the quadratic and linear momentum terms in the dispersion relation, respectively.  This redundancy is consistent with the perturbative spinor-field redefinitions discussed in Ref.~\cite{Colladay2002}, according to which a coefficient such as $b^{\mu}$ multiplying the identity matrix in the generalized Dirac operator may be perturbatively eliminated in favor of the remaining anisotropic coefficients. Since the purpose of the present work is to characterize the independent geometric deformations of Dirac cones, we adopt the convenient parametrization $b^{\mu} = 0$ throughout this paper. This choice, however, does not imply that the corresponding operator is physically irrelevant in more general contexts, where it may lead to observable effects beyond the geometric properties of the Dirac cones. Such situations are beyond the scope of the present work.

 Now, Eq.~(\ref{dispersionRel}) can be written in component form as 
\begin{equation}
\left(1+\Upsilon^{00}\right)\left(p^{0}\right)^{2}-2\left(\Upsilon^{0i}p^{i}+\Theta^{0}\right)p^{0}=\vec{p}^{2}-\Upsilon^{ij}p^{i}p^{j}-v_{F}^{2}\left[\left(d^{0}\right)^2 - \vec{d}^2 - m^2\right]-2\Theta^{i}p^{i},
\label{onlyComponents}
\end{equation}
where we used 
\begin{equation}p^2 = \eta_{\mu\nu} p^{\mu} p^{\nu} = \left(p^{0}\right)^2 - \vec{p}^2,
\end{equation}
\begin{equation}d^2 = \eta_{\mu\nu} d^{\mu} d^{\nu} = \left(d^{0}\right)^2 - \vec{d}^2,
\end{equation}
\begin{equation}
\Upsilon^{\mu\nu}p_{\mu}p_{\nu} = \eta_{\mu\sigma}\eta_{\nu\rho}\Upsilon^{\mu\nu}p^{\sigma}p^{\rho} = \Upsilon^{00}\left(p^{0}\right)^{2}-2\Upsilon^{0i}p^{0}p^{i}+\Upsilon^{ij}p^{i}p^{j}
\end{equation}
and
\begin{equation}
\Theta^{\mu}p_{\mu} = \eta_{\mu\sigma}\Theta^{\mu}p^{\sigma } = \Theta^{0}p^{0}-\Theta^{i}p^{i}.
\end{equation}
Multiplying \eqref{onlyComponents} by $\left(1+\Upsilon^{00}\right)^{-1}$ and  completing the square, we obtain
\begin{equation}
p^{0}=v_{0}+\left(\vec{p}\cdot\vec{w}\right)\pm\sqrt{\mathcal{A}_{ij}p^{i}p^{j}-2v_{i}p^{i}+\mu^{2}},
\label{CompReldispersion}
\end{equation}
where
\begin{eqnarray}
v_{0}&=&\frac{v_{F}\Theta^{0}}{\left(1+\Upsilon^{00}\right)},\label{definitionsParameters0}\\  
w_{i}&=&\frac{\Upsilon^{0i}}{\left(1+\Upsilon^{00}\right)},\label{wdef}\\
\mathcal{A}_{ij}&=&\frac{\delta_{ij}-\Upsilon^{ij}}{\left(1+\Upsilon^{00}\right)}+w_{i}w_{j},\label{Amatrix}\\
v_{i}&=&\frac{v_{F}\Theta^{i}}{\left(1+\Upsilon^{00}\right)}-v_{0}w_{i},\label{Vmatrix}\\
\mu^{2}&=&\frac{\vec{d}^{2}-\left(d^{0}\right)^{2} + m^2}{\left(1+\Upsilon^{00}\right)}v_{F}^{2}+\left(v_{0}\right)^{2} \label{muSquared}.
\label{definitionsParameters}
\end{eqnarray}
Here, \(i,j=1,2\), \(\vec{w} \equiv (w_{1}, w_{2})\) is a constant vector, and \(\delta_{ij}\) is the Kronecker delta. At this stage, the result is fully general.  We will use the notation $1 \equiv  x$ and  $2 \equiv   y$ throughout the remainder of this paper. Accordingly, $\vec{d} = (d^{1},d^{2}) = (d_{x},d_{y})$ and $\vec{p} = (p^{1},p^{2}) = (p_{x},p_{y})$.

Let us now discuss the approximation in which the effects produced by the parameters in the Lagrangian density can be treated perturbatively. This approximation is justified by considering a natural energy scale $E_{c}$, assumed to be characteristic of the two-dimensional Dirac material under consideration. In the case of the parameters \(d^{\mu}\) and \(m\), which have dimensions of mass, a perturbative interpretation of the effects requires \(\mid d^{\mu} \mid v_{F}^{2} << E_{c}\) and \(m v_{F}^2 << E_{c}\). On the other hand, since \(a^{\mu}_{\,\,\nu}\) and \(b^{\mu}\) are dimensionless, it is sufficient to require \(\mid a^{\mu\nu}\mid, \mid b^{\mu}\mid << 1\). Moreover, from the perspective of effective field theory, the smallness of the free parameters relative to \(E_{c}\) ensures that the modified Dirac equation,  given by Eq.~(\ref{modEq}), represents a controlled expansion around the isotropic, massless limit. The choice of the scale \(E_{c}\) will be discussed in Sec.~\ref{validation}, where the model is applied to real materials.

Since the matrix $\mathcal{A}_{ij}$ is symmetric and, in the   perturbative regime considered here,  positive definite, the argument of the square root
\begin{equation}
\mathcal{R}(p_{x} , p_{y}) = \mathcal{A}_{ij}p^{i}p^{j}-2v_{i}p^{i}+\mu^{2}
\label{rootArgument} 
\end{equation} 
has a global minimum given by $\mathcal{R}_{min} = v_{F}^{2}\frac{m^2}{1 + \Upsilon^{00}}$ (see the Appendix~\ref{proofminimum} for the proof). In fact, by completing squares, Eq.~\eqref{rootArgument} can be written as
\begin{equation}
\mathcal{R}(p_{x},p_{y})=\left({\bf P}-{\bf P}_{0}\right)^{T}\mathcal{A}\left({\bf P}-{\bf P}_{0}\right)-\mathcal{V}^{T}\mathcal{A}^{-1}\mathcal{V}+\mu^{2}
\label{dispersionSquared}.
\end{equation} 
Here, we defined
\begin{equation}{\bf P}=\left(\begin{array}{c}
p_{x}\\
p_{y}
\end{array}\right)\end{equation}
and
\begin{equation}{\bf P}_{0}=\mathcal{A}^{-1}\mathcal{V},\label{deg_point}\end{equation}
with
\begin{equation}
\mathcal{V}=\left(\begin{array}{c}
v_{x}\\
v_{y}
\end{array}\right).
\end{equation}
 $\mathcal{A}^{-1}$ is the inverse of the matrix $\mathcal{A}$. The matrices $\mathcal{A}$ and $\mathcal{V}$ are defined by the components in Eqs.~\eqref{Amatrix} and \eqref{Vmatrix}, respectively. As shown in the Appendix~\ref{proofminimum},
\begin{equation}
\mathcal{V}^{T}\mathcal{A}^{-1}\mathcal{V} = \frac{\vec{d}^{2}-\left(d^{0}\right)^{2}}{\left(1+\Upsilon^{00}\right)}v_{F}^{2}+\left(v_{0}\right)^{2}.
\end{equation}
So, using this expression in Eq.~\eqref{dispersionSquared}, we have
\begin{equation}
\mathcal{R}(p_{x},p_{y})=\left({\bf P}-{\bf P}_{0}\right)^{T}\mathcal{A}\left({\bf P}-{\bf P}_{0}\right) + v_{F}^{2} \frac{m^2}{\left(1+\Upsilon^{00}\right)}.
\label{argumentRootSquaresm}
\end{equation}
Thus, for $m=0$, the dispersion relation in Eq.~\eqref{CompReldispersion} describes Dirac cones with a single degeneracy point at ${\bf P}_{0}$, given by Eq.~\eqref{deg_point}. For $m\neq0$, the system exhibits a mass gap centered at this point. In other words, the first case  describes the usual massless semimetallic phase, whereas the second  a semiconducting phase. This result shows that the anisotropic parameters $a^{\mu\nu}$, $b^{\mu} $  and even $d^{\mu}$  do not break the parity symmetry, which is broken exclusively  by the parameter $m$.

In order to simplify the analysis of the physical effects discussed in the next section, it is useful to make explicit the form of the dispersion relation in Eq.~\eqref{CompReldispersion} by retaining only the dominant contributions of the parameters $a^{\mu\nu}$ and $d^{\mu}$ to $\mathcal{A}$ and ${\bf P}_{0}$. In this approximation, $\Upsilon_{\mu\nu}$ and $\Theta_\mu$, defined in Eqs.~\eqref{upsilon} and \eqref{theta}, are evaluated by retaining only the first-order contributions in the parameters $a^{\mu\nu}$ and $d^\mu$. Thus, from Eqs.~\eqref{Amatrix} and \eqref{Vmatrix} we obtain
\begin{equation}
\mathcal{A}\approx\left(\begin{array}{cc}
1-a^{xx} & a^{xy}\\
a^{xy} & 1-a^{yy}
\end{array}\right)
\label{approxmatrix}
\end{equation} 
and 
\begin{equation}
{\bf P}_{0}  \approx\left(\begin{array}{c}
v_{F} d_{x}\\
v_{F}  d_{y}
\end{array}\right).
\label{approxP0}
\end{equation}
We have used that  $\det\mathcal{A}\approx1$ and $\left(1+\Upsilon^{00}\right)^{-1} \approx 1- a^{00}$. For notational simplicity, the factor of 2 in Eq.~\eqref{upsilon} has been absorbed into the definition of \(a^{\mu\nu}\), i.e., \(2a^{\mu\nu} \rightarrow a^{\mu\nu}\). In addition, within this approximation, the component \(a^{00}\) has been absorbed through the redefinitions \(a^{00}+a^{xx}\rightarrow a^{xx}\) and \(a^{00}+a^{yy}\rightarrow a^{yy}\).

Using the approximations defined by Eqs.~\eqref{approxmatrix} and \eqref{approxP0} in Eq.~\eqref{argumentRootSquaresm}, and retaining only the leading contribution in $m^2$, we obtain 
\begin{equation}
\mathcal{R}(p_{x},p_{y}) = \left(1-a^{xx}\right)\left(p_{x}-v_{x}\right)^{2}+\left(1-a^{yy}\right)\left(p_{y}-v_{y}\right)^{2}+2a^{xy}\left(p_{x}-v_{x}\right)\left(p_{y}-v_{y}\right)+v_{F}^{2}m^{2}.
\label{arg_approx}
\end{equation}
In particular, corrections of order $m^2\Upsilon^{00}$ are neglected consistently with the perturbative approximation. Eq.~\eqref{arg_approx} is a convenient  expression for Eq.~\eqref{rootArgument}, which,  when substituted into Eq.~\eqref{CompReldispersion}, yields
\begin{equation}
\tilde{E}=\tilde{E}_{0}+\tilde{p}_{x}a^{0x}+\tilde{p}_{y}a^{0y}\pm\sqrt{\left(1-a^{xx}\right)\left(\tilde{p}_{x}-\tilde{d}_{x}\right)^{2}+\left(1-a^{yy}\right)\left(\tilde{p}_{y}-\tilde{d}_{y}\right)^{2}+2a^{xy}\left(\tilde{p}_{x}-\tilde{d}_{x}\right)\left(\tilde{p}_{y}-\tilde{d}_{y}\right)+\tilde{m}^{2}}\, ,
\label{adim_dispertion}
\end{equation}
where
\begin{eqnarray}
\vec{\tilde{p}}&=&\frac{v_{F}}{E_{c}}\vec{p},\\
\vec{\tilde{d}}&=&\frac{v_{F}^{2}}{E_{c}}\vec{d},\\
\tilde{m}&=&\frac{mv_{F}^{2}}{E_{c}},\\
\tilde{E}_{0}&=&\frac{E_{0}}{E_{c}}=\frac{v_{F}^{2}}{E_{c}}d^{0},\\
\tilde{E}&=&\frac{E}{E_{c}}.
\end{eqnarray}
Here, the physical quantities are expressed in terms of dimensionless quantities relative to the effective energy scale \(E_{c}\), which serves as a cutoff for the model. As expected, when all the parameters in Eq.~\eqref{adim_dispertion} vanish, we recover the usual isotropic massless Dirac cones, which exhibit the conventional zero-gap semimetal behavior around the Dirac points \(K\) and \(K'\) (valleys) observed, for instance, in graphene~\cite{NOVOSELOV}. 
\section{Anisotropic Effects from the Extended Dirac Equation}
\label{AnisotropicEffects}
 In this section, we analyze the physical consequences of the anisotropic extension of the Dirac equation introduced above. The numerical analysis reveals that the Lorentz-violating coefficients play distinct and complementary roles in determining the geometry of the Dirac spectrum. Notably, the model captures a class of modifications that emerge from the two-dimensional nature of the system, rather than from specific properties of particular materials. Our focus is on the effects produced individually by the parameters $a^{\mu\nu}$ and $d^{\mu}$ on the Dirac cones, for both cases $m = 0$ and $m \ne 0$. The resulting effects can be associated with experimentally observed features in several 2D Dirac materials, as will be discussed in the next section.
\subsection{Effects of the \(d^\mu\) parameter}
Let us begin by analyzing the effects produced exclusively by the parameter \(d^\mu\). It is interesting to note that the Lagrangian density in Eq.~(\ref{lagrangian})  can be written as
\begin{equation}
\mathcal{L}=i\hbar v_{F}\bar{\psi}\Gamma^{\mu}D_{\mu}\psi _{\mu}\psi- v_{F}^{2}\bar{\psi} M \psi,
\label{lagrangianCov}
\end{equation}
where \(D_{\mu} = \partial_{\mu}+\frac{i}{\hbar} A_\mu\) plays the role of a covariant derivative and \(A_\mu = v_{F} d_{\mu}\) acts as an effective gauge field. Thus, \( d_\mu \) behaves in this two-dimensional model as a fixed background field without intrinsic dynamics, similarly to what occurs in Lorentz-violating effective models. From the experimental perspective, pseudo-gauge fields have been observed through strain engineering in graphene~\cite{LEVY}. Furthermore, tight-binding models of 8-Pmmn borophene indicate the emergence of pseudo-magnetic vector potentials and scalar potentials induced by strain~\cite{Zabolotskiy}. In fact, strain-induced gauge fields are by now well established in the literature~\cite{VOZMEDIANO}. Here, we show that they emerge naturally as a general feature of our two-dimensional model for Dirac materials based on Eq.~(\ref{modEq}). It should be noted that in the model we are considering the pseudogauge field does not induce  electromagnetic-like effects, since $d^{\mu}$ is constant.

Let us now consider the dispersion relation in Eq.~\eqref{adim_dispertion} assuming that only the parameters  $d^0$ and $\vec{d}$ are nonzero. It reproduces well-established effects associated with strain-induced  pseudoscalar and pseudovector potentials in graphene. Indeed, the components of the pseudo-gauge field \(A_{\mu}\) defined in Eq.~(\ref{lagrangianCov}) are precisely those appearing in Eq.~(\ref{adim_dispertion}). Since we consider constant \(d^{\mu}\), corresponding to uniform strain, the model does not open a gap between the valence and conduction bands, but instead produces only a shift of the conical points~\cite{Katsnelson}.

\begin{figure}[htbp]
\centering
\includegraphics[width=0.35\textwidth]{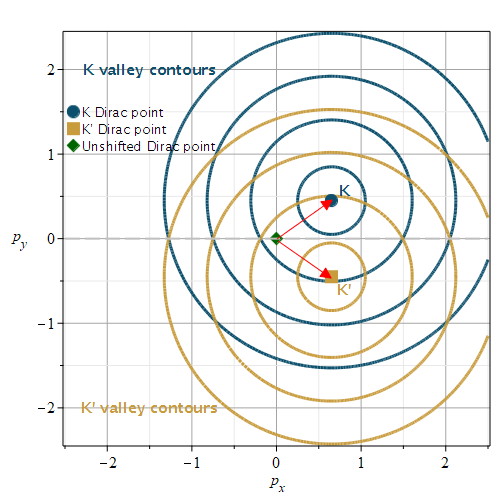}
\caption{\small
Parity-related shifts of the Dirac points: the constant-energy contours around the Dirac valleys $K$ and $K'$, related by the parity transformation $(p_x,p_y)\rightarrow(p_x,-p_y)$.
}
\label{fig:all_figures0}
\end{figure}

\begin{figure}[htbp]
\centering
\begin{minipage}{0.32\textwidth}
\centering
\includegraphics[width=\linewidth]{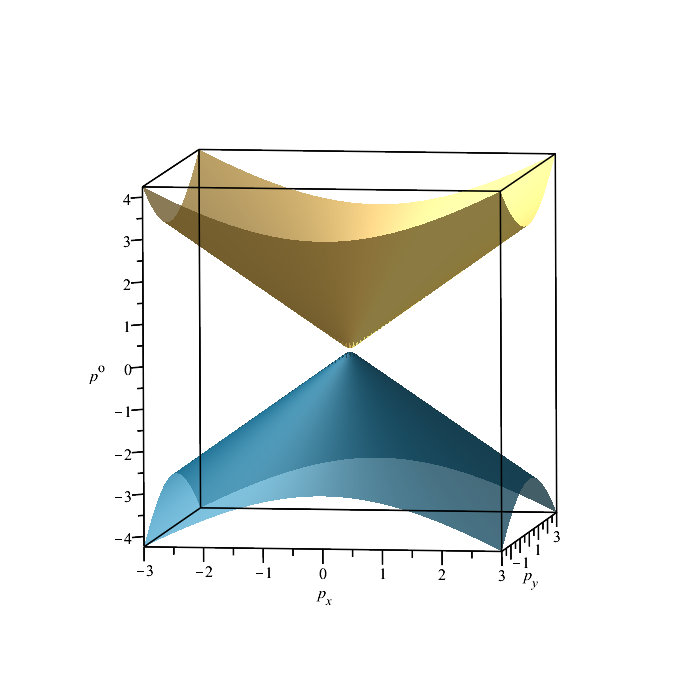}
\textbf{(a)} Pristine Dirac cone\\($m=0$).
\end{minipage}
\begin{minipage}{0.32\textwidth}
\centering
\includegraphics[width=\linewidth]{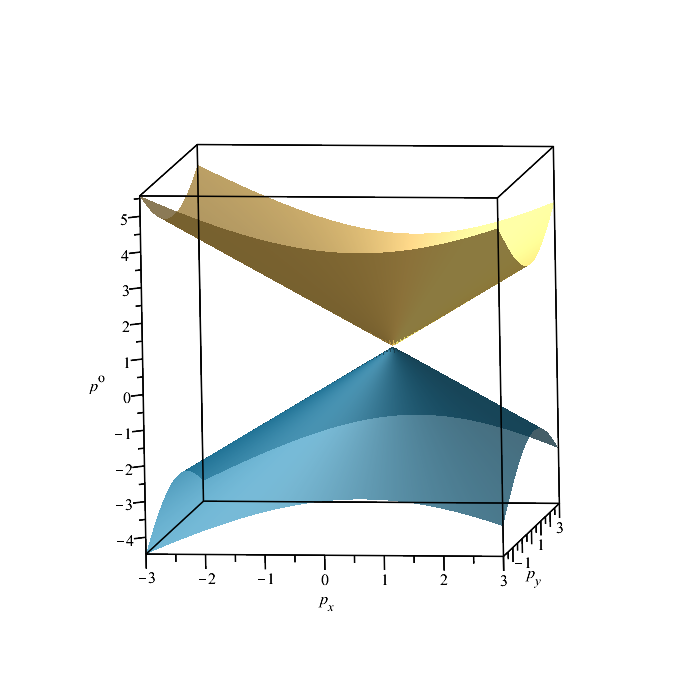}
\textbf{(b)} Dispersion relation\\($m=0$).
\end{minipage}
\begin{minipage}{0.32\textwidth}
\centering
\includegraphics[width=\linewidth]{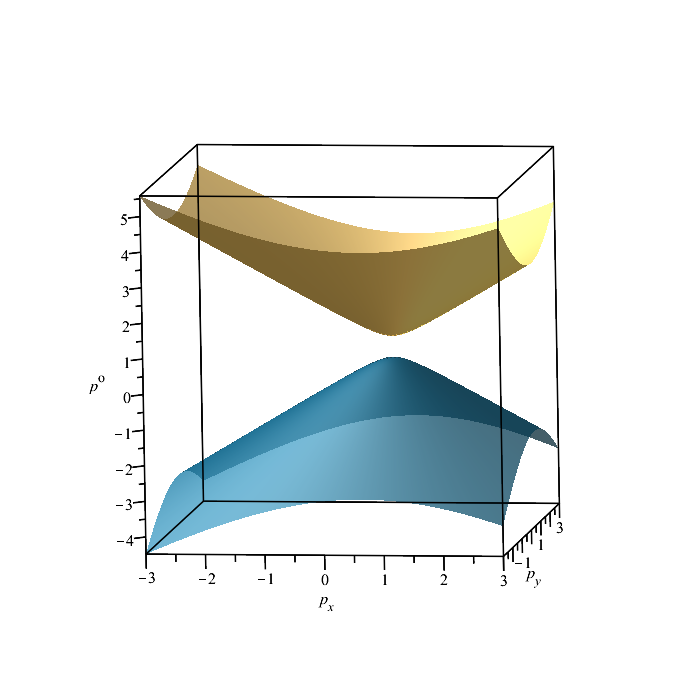}
\textbf{(c)} Dispersion relation\\($m\neq0$).
\end{minipage}
\vspace{0.1cm}
\begin{minipage}{0.30\textwidth}
\centering
\includegraphics[width=\linewidth]{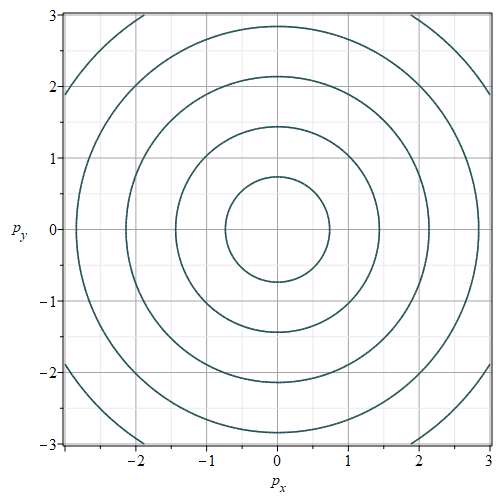}
\textbf{(d)} Contour plot\\(Pristine Dirac cones).
\end{minipage}
\begin{minipage}{0.30\textwidth}
\centering
\includegraphics[width=\linewidth]{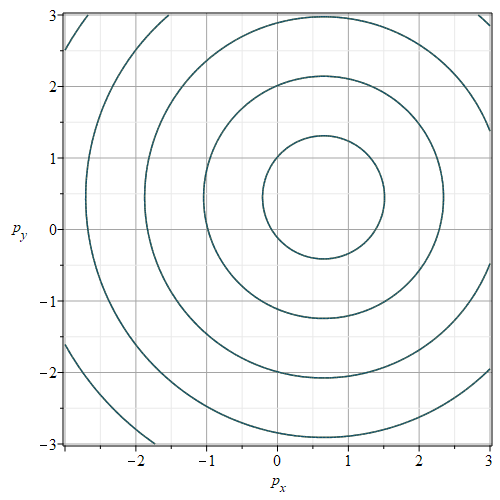}
\textbf{(e)} Contour plot\\($m=0$).
\end{minipage}
\begin{minipage}{0.30\textwidth}
\centering
\includegraphics[width=\linewidth]{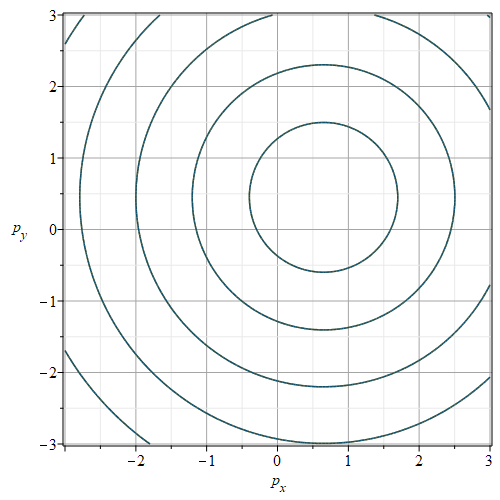}
\textbf{(f)} Contour plot\\($m\neq0$).
\end{minipage}

\caption{\small
Effects generated exclusively by the vector $d^\mu$, with all the other anisotropic parameters $a^{\mu\nu}$ set to zero. The parameter $d^\mu$ preserves the conical dispersion and acts only by shifting the Dirac cones in momentum and energy space. While the spatial components $d_x$ and $d_y$ shift the Dirac point along the momentum direction,} the temporal component $d^0$ introduces an overall energy displacement. The plots were generated in dimensionless units ($v_F=1$) with $d^0=0.55$, $d_x=0.65$, and $d_y=0.45$.

\label{fig:all_figures1}
\end{figure}

Fig.~\ref{fig:all_figures0} shows the constant-energy contours associated with the two Dirac points \(K\) and \(K'\), which are related by the parity transformation \((d_x,d_y)\rightarrow(d_x,-d_y)\) in \((2+1)\) dimensions, leading to parity-related shifts of the Dirac points in momentum space. The figure makes explicit that the parameter \(\vec d\) produces only a displacement of the Dirac cones while preserving their conical structure. In Fig.~\ref{fig:all_figures1}, it can be seen that the vector coefficient $d^\mu$ does not alter the conical character of the dispersion relation. Instead, its temporal component produces a rigid energy shift, whereas the spatial components displace the Dirac points in momentum space without modifying the local cone geometry. Consequently, the constant-energy contours preserve their circular shape, indicating that the Fermi velocity remains isotropic.
  When $m\neq0$, a mass gap is generated between the valence and conduction bands.
\FloatBarrier
\subsection{Effects of the $a^{\mu\nu}$ parameter}
\begin{figure}[htbp]
\centering
\begin{minipage}{0.3\textwidth}
\centering
\includegraphics[width=\linewidth]{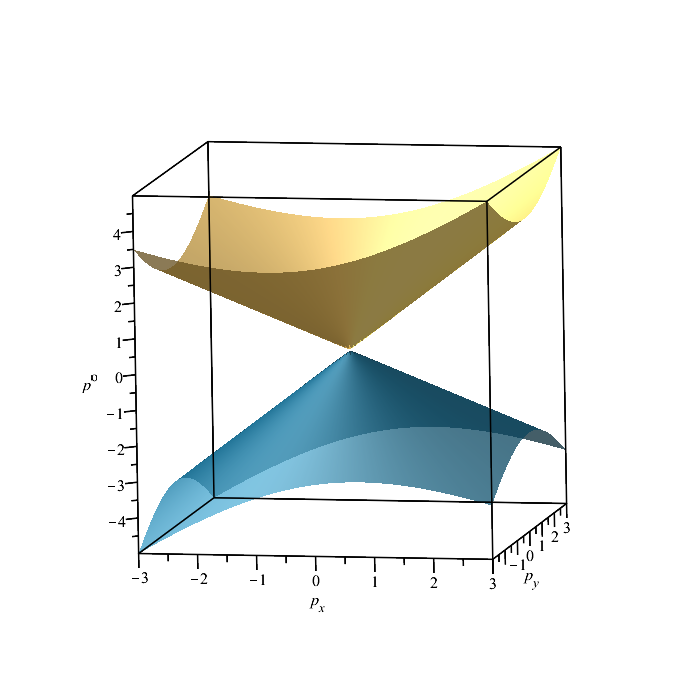}
{\footnotesize\textbf{(a)} Type-I tilted Dirac cone\\ ($a^{0x}=0.25$, $a^{xx}=0$).}
\end{minipage}
\begin{minipage}{0.3\textwidth}
\centering
\includegraphics[width=\linewidth]{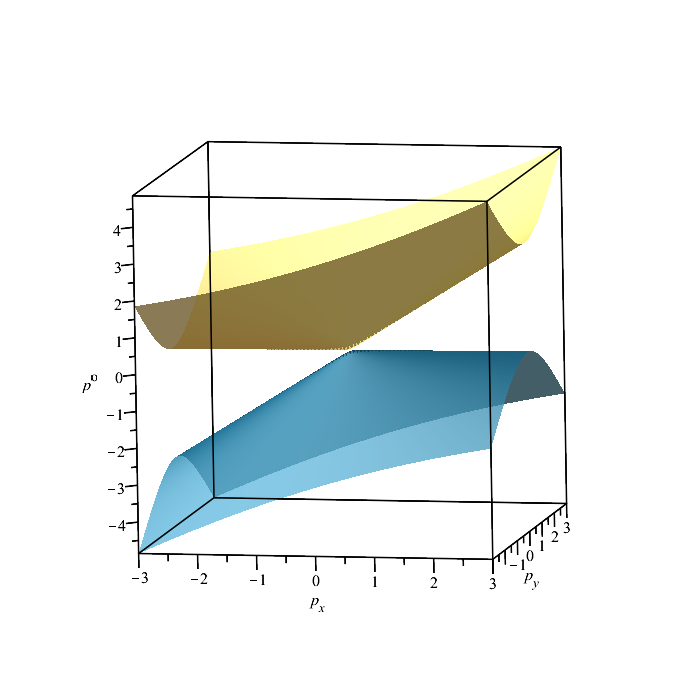}
{\footnotesize\textbf{(b)} Type-III tilted Dirac cone\\ ($a^{0x}=0.50$, $a^{xx}=0.75$).}
\end{minipage}
\begin{minipage}{0.3\textwidth}
\centering
\includegraphics[width=\linewidth]{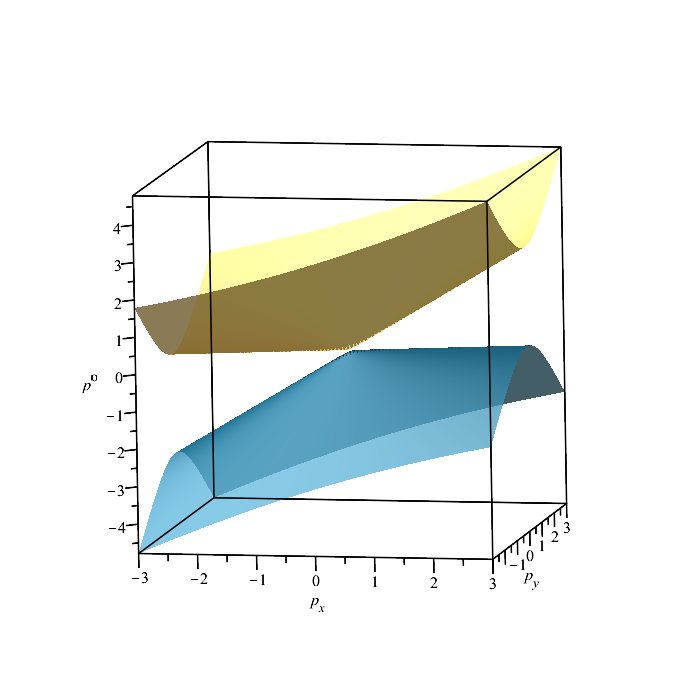}
{\footnotesize\textbf{(c)} Type-II tilted Dirac cone\\ ($a^{0x}=0.50$, $a^{xx}=0.80$).}
\end{minipage}
\vspace{0.1cm}
\begin{minipage}{0.26\textwidth}
\centering
\includegraphics[width=\linewidth]{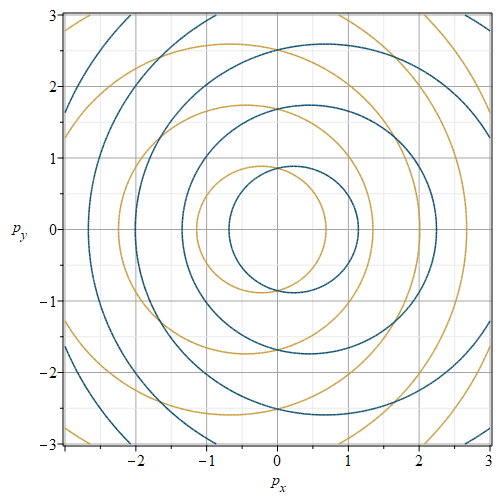}
{\footnotesize\textbf{(d)} Contour plot (Type-I).}
\end{minipage}
\begin{minipage}{0.26\textwidth}
\centering
\includegraphics[width=\linewidth]{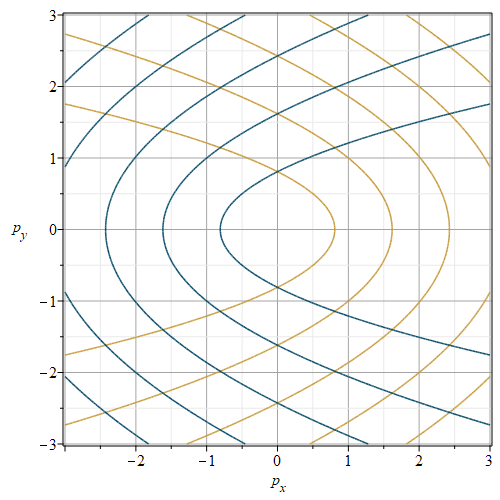}
{\footnotesize\textbf{(e)} Contour plot (Type-III).}
\end{minipage}
\begin{minipage}{0.26\textwidth}
\centering
\includegraphics[width=\linewidth]{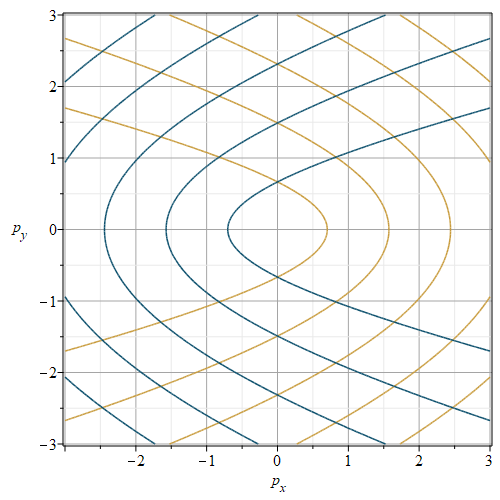}
{\footnotesize\textbf{(f)} Contour plot (Type-II).}
\end{minipage}
\vspace{0.1cm}
\begin{minipage}{0.26\textwidth}
\centering
\includegraphics[width=\linewidth]{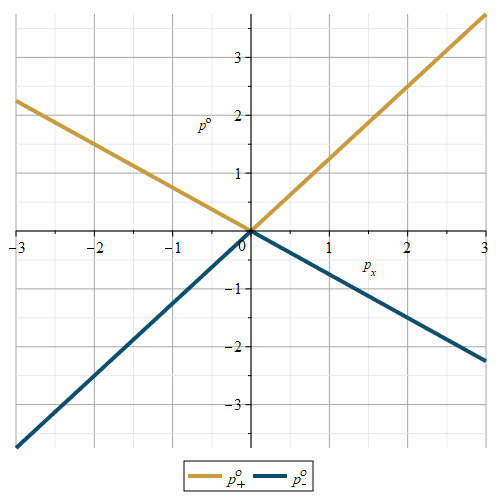}
{\footnotesize\textbf{(g)} Energy cut along $p_y=0$ (Type-I).}
\end{minipage}
\begin{minipage}{0.26\textwidth}
\centering
\includegraphics[width=\linewidth]{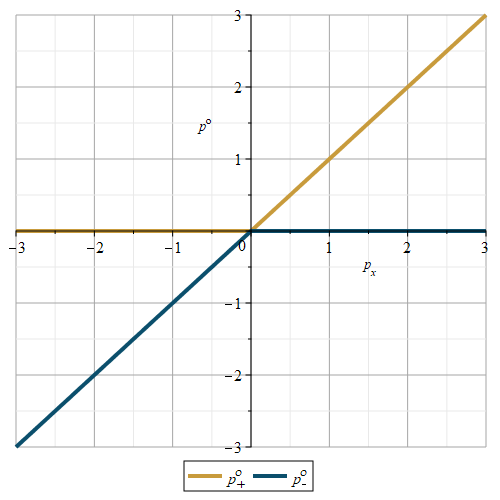}
{\footnotesize\textbf{(h)} Energy cut along $p_y=0$ (Type-III).}
\end{minipage}
\begin{minipage}{0.26\textwidth}
\centering
\includegraphics[width=\linewidth]{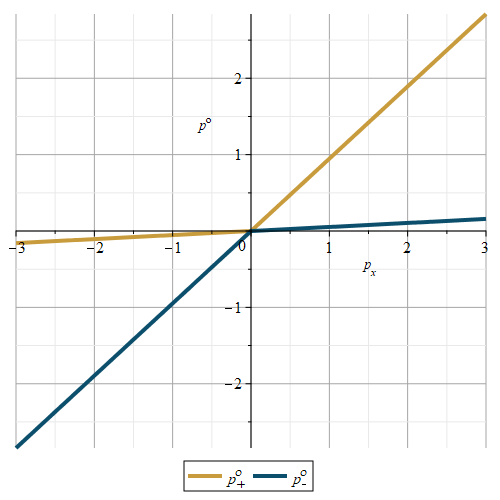}
{\footnotesize\textbf{(i)} Energy cut along $p_y=0$ (Type-II).}
\end{minipage}
\vspace{0.1cm}
\begin{minipage}{0.26\textwidth}
\centering
\includegraphics[width=\linewidth]{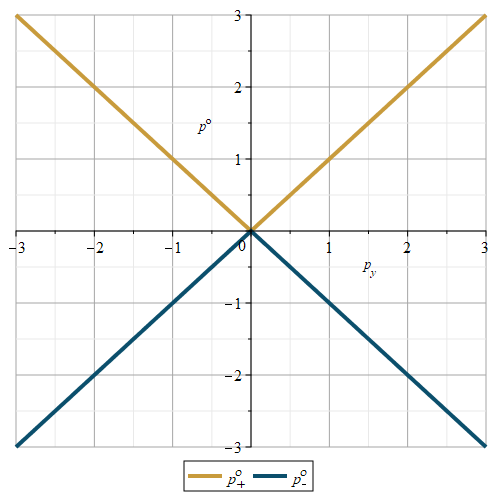}
{\footnotesize\textbf{(j)} Energy cut along $p_x=0$ (Type-I).}
\end{minipage}
\begin{minipage}{0.26\textwidth}
\centering
\includegraphics[width=\linewidth]{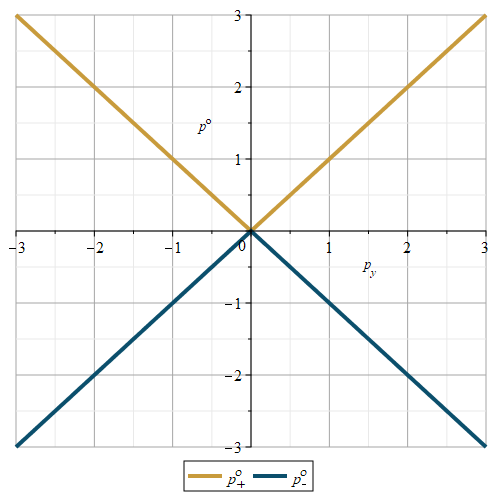}
{\footnotesize\textbf{(k)} Energy cut along $p_x=0$ (Type-III).}
\end{minipage}
\begin{minipage}{0.26\textwidth}
\centering
\includegraphics[width=\linewidth]{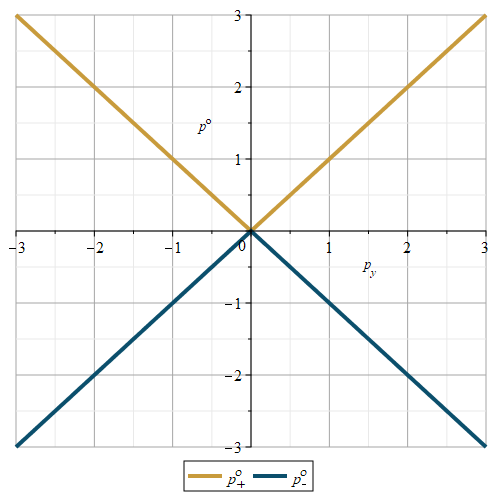}
{\footnotesize\textbf{(l)} Energy cut along $p_x=0$ (Type-II).}
\end{minipage}
\caption{\small
Evolution of the Dirac-cone geometry induced by the anisotropic parameters
$a^{0x}$ and $a^{xx}$, with all other coefficients set to zero,
$a^{00}=a^{0y}=a^{yy}=a^{xy}\equiv 0$. (a)--(f): the effective tilt and anisotropy are governed by the coefficients $w_{0x}$ and $\mathcal{A}_{ij}$ entering the dispersion relation, Eq.~\ref{adim_dispertion}.
The sequence shows the continuous transition from a conventional Type-I Dirac cone, through the critical Type-III state, to an overtilted Type-II Dirac cone. Panels (g)--(l) present the corresponding energy cuts along $p_y=0$ and $p_x=0$. All quantities are given in dimensionless units with $v_F=1$.}
\label{fig:all_figures}
\end{figure}

Qualitatively different behaviors emerge from the coefficients $a^{\mu\nu}$. We start by considering \(d^{\mu} = 0\) and \(m=0\) in Eq.~(\ref{adim_dispertion}). The effects of the coefficient $a^{0x}$, combined with different values of $a^{xx}$, as shown in Fig.~\ref{fig:all_figures}, induce a progressive tilting of the Dirac cone while preserving its gapless nature. As their magnitudes increase, the system evolves continuously from a conventional Type-I Dirac cone through the critical Type-III configuration and finally into the overtilted Type-II regime. The corresponding energy cuts clearly reveal this transition, demonstrating how the tilt modifies the relative slopes of the conduction and valence bands and ultimately gives rise to open constant-energy contours, which are the characteristic signature of Type-II Dirac fermions. These are well-established features of two-dimensional Dirac materials and have been reported in systems exhibiting elliptical Dirac cones~\cite{IUROV}, as well as tilted Type-I and Type-II Dirac cones and the critical Type-III transition~\cite{GOERBIG2,MILIĆEVIĆ}.

\begin{figure}[htbp]
\centering
\begin{minipage}{0.3\textwidth}
\centering
\includegraphics[width=\linewidth]{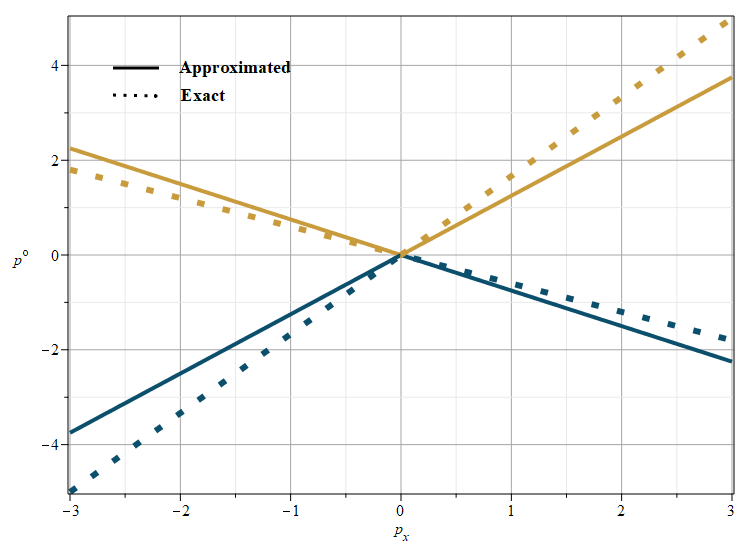}
{\footnotesize\textbf{(a)} Type-I tilted Dirac cone\\ ($a^{0x}=0.25$, $a^{xx}=0$, \\$\tilde{w}_0<1$).}
\end{minipage}
\begin{minipage}{0.3\textwidth}
\centering
\includegraphics[width=\linewidth]{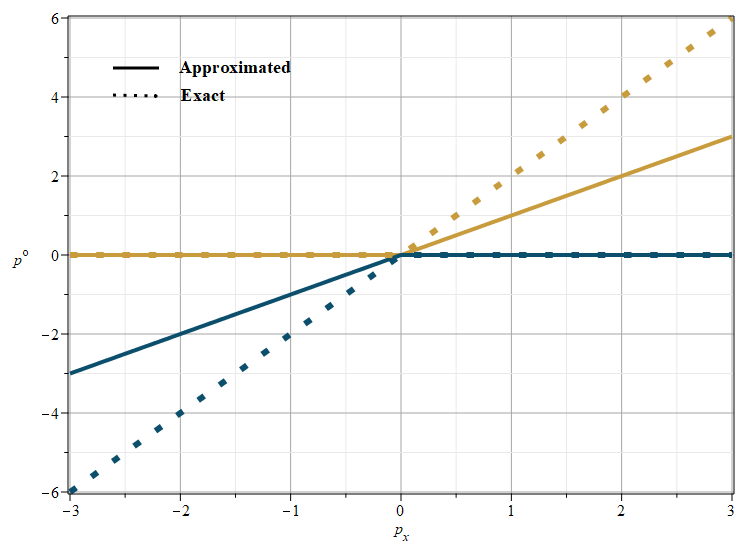}
{\footnotesize\textbf{(b)} Type-III tilted Dirac cone\\ $(a^{0x}=0.50$, $a_{\rm approx}^{xx}=0.75$, $a_{\rm exact}^{xx}=0.50$, $\tilde{w}_0=1$).}
\end{minipage}
\begin{minipage}{0.3\textwidth}
\centering
\includegraphics[width=\linewidth]{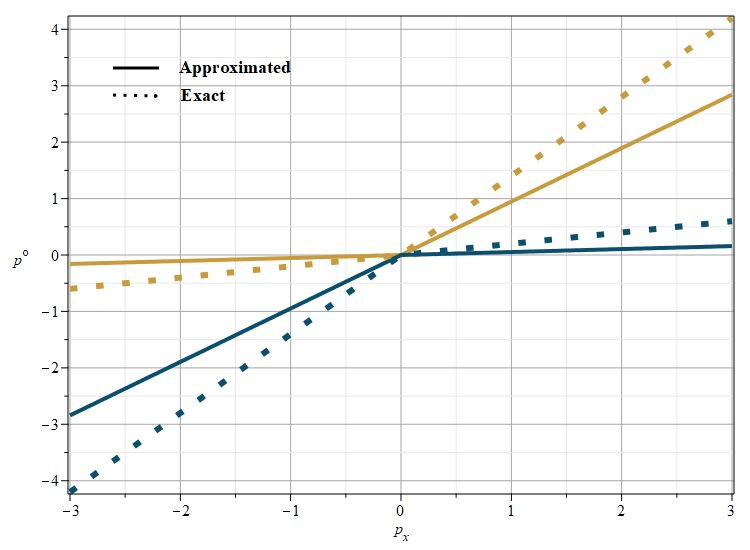}
{\footnotesize\textbf{(c)} Type-II tilted Dirac cone\\ ($a^{0x}=0.50$, $a^{xx}=0.80$,\\ $\tilde{w}_0>1$).}
\end{minipage}
\caption{\small Comparison between the approximate (solid lines) and exact (dotted lines) energy dispersions along the cut $p_y=0$ for the three types of tilted Dirac cones.
(a) Type-I Dirac cone, obtained for $a^{0x}=0.25$ and $a^{xx}=0$ in both descriptions, satisfying $\tilde{w}_0<1$.
(b) Critical Type-III Dirac cone, where the approximate and exact theories require different values of the anisotropy coefficient to satisfy the critical condition, namely $a^{xx}_{\mathrm{approx}}=0.75$ and $a^{xx}_{\mathrm{exact}}=0.50$, respectively, with $a^{0x}=0.50$ and $\tilde{w}_0=1$.
(c) Type-II Dirac cone, obtained for $a^{0x}=0.50$, $a^{xx}=0.80$, and $\tilde{w}_0>1$.
The comparison illustrates that the approximate dispersion reproduces the qualitative evolution of the cone geometry but exhibits quantitative deviations from the exact solution as the anisotropic parameters increase, becoming particularly pronounced in the Type-II regime.
All quantities are expressed in dimensionless units with $v_F=1$.
}
\label{comp}
\end{figure}

Figure~\ref{comp} compares the energy cuts along $p_y=0$ obtained from the approximate and exact dispersion relations for the three classes of tilted Dirac cones, given by Eqs.~\eqref{CompReldispersion} and \eqref{adim_dispertion}, respectively. The approximate solutions are represented by solid lines, whereas the exact results are shown as dotted lines. For the Type-I regime, both descriptions produce nearly identical dispersions, indicating that the approximate theory accurately reproduces the low-anisotropy limit. As the anisotropic parameters increase, quantitative differences become progressively more pronounced, particularly in the Type-II regime, where the slopes of the bands and the effective tilt differ noticeably.

Despite these quantitative deviations, the approximate theory successfully captures all qualitative features of the exact dispersion. In particular, it correctly describes the continuous evolution from Type-I to Type-III and finally to the overtilted Type-II Dirac cone, preserving the characteristic band topology associated with each regime. The critical Type-III configuration is also recovered, although it requires a different numerical value of the anisotropy coefficient $a^{xx}$ than that obtained from the exact theory.

\begin{figure}[htbp]
\centering
\begin{minipage}{0.32\textwidth}
\centering
\includegraphics[width=\linewidth]{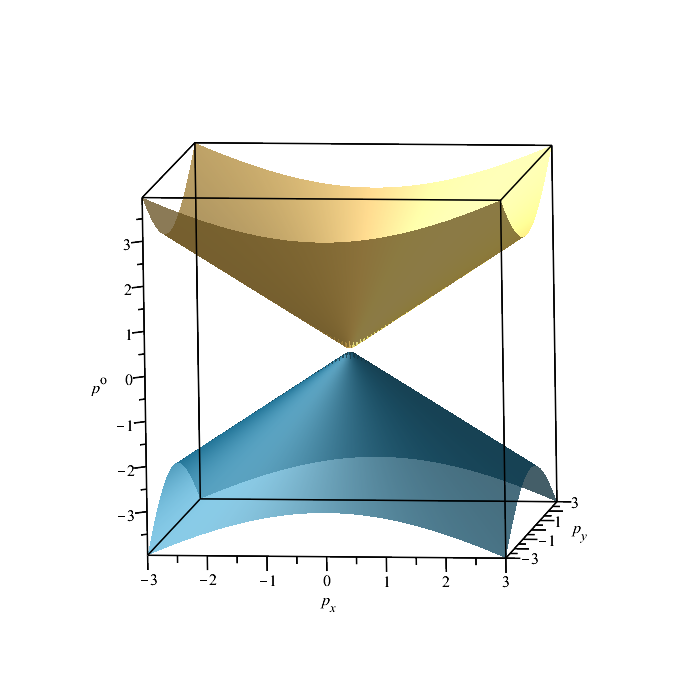}
{\footnotesize\textbf{(a)} Dispersion relation ($a^{xx}=0.25$).}
\end{minipage}
\begin{minipage}{0.32\textwidth}
\centering
\includegraphics[width=\linewidth]{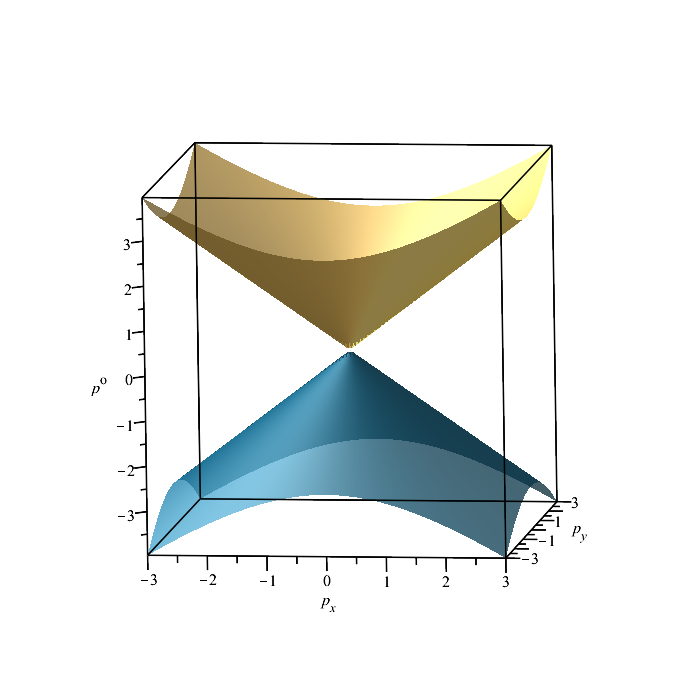}
{\footnotesize\textbf{(b)} Dispersion relation ($a^{yy}=0.25$).}
\end{minipage}
\begin{minipage}{0.32\textwidth}
\centering
\includegraphics[width=\linewidth]{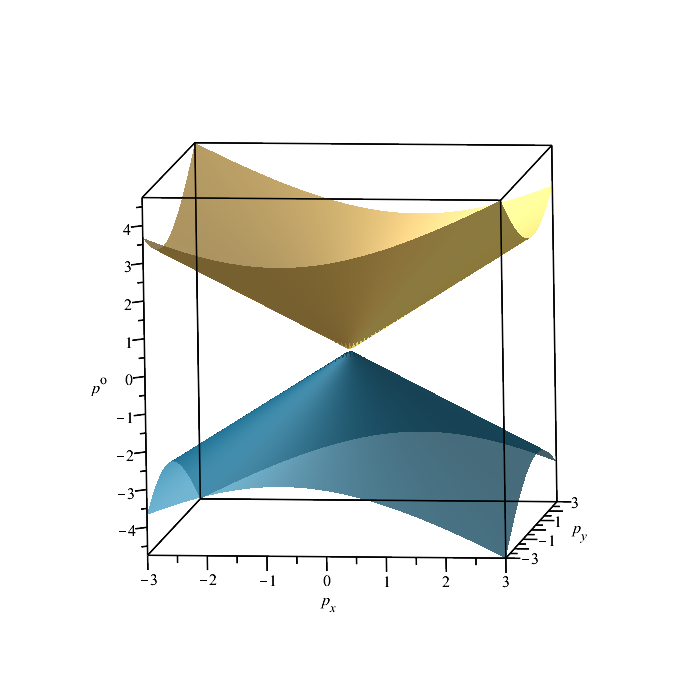}
{\footnotesize\textbf{(c)} Dispersion relation ($a^{xy}=0.25$).}
\end{minipage}

\vspace{0.25cm}

\begin{minipage}{0.30\textwidth}
\centering
\includegraphics[width=\linewidth]{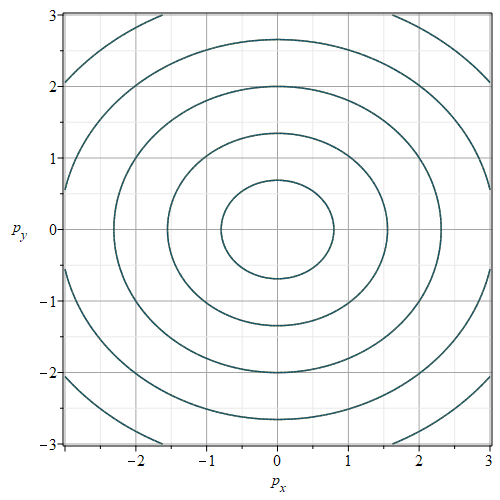}
{\footnotesize\textbf{(d)} Contour plots ($a^{xx}=0.25$).}
\end{minipage}
\begin{minipage}{0.30\textwidth}
\centering
\includegraphics[width=\linewidth]{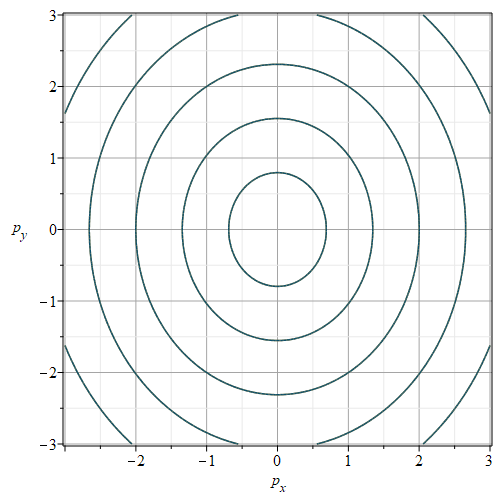}
{\footnotesize\textbf{(e)} Contour plots ($a^{yy}=0.25$).}
\end{minipage}
\begin{minipage}{0.30\textwidth}
\centering
\includegraphics[width=\linewidth]{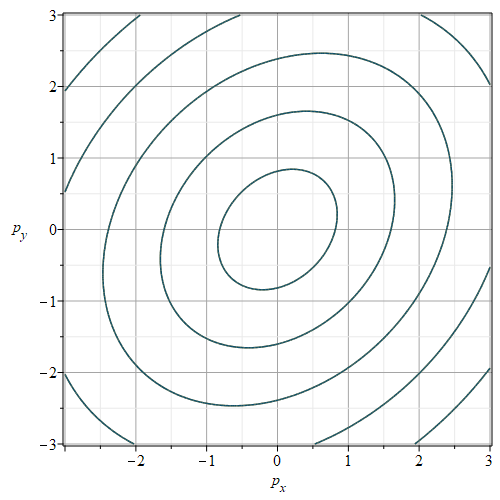}
{\footnotesize\textbf{(f)} Contour plots ($a^{xy}=0.25$).}
\end{minipage}

\vspace{0.25cm}

\begin{minipage}{0.30\textwidth}
\centering
\includegraphics[width=\linewidth]{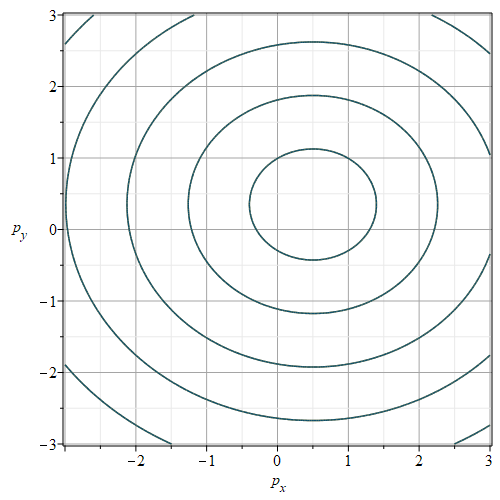}
{\footnotesize\textbf{(g)} Contour plots ($a^{xx}=0.25$, $d_x=0.50$, $d_y=0.35$).}
\end{minipage}
\begin{minipage}{0.30\textwidth}
\centering
\includegraphics[width=\linewidth]{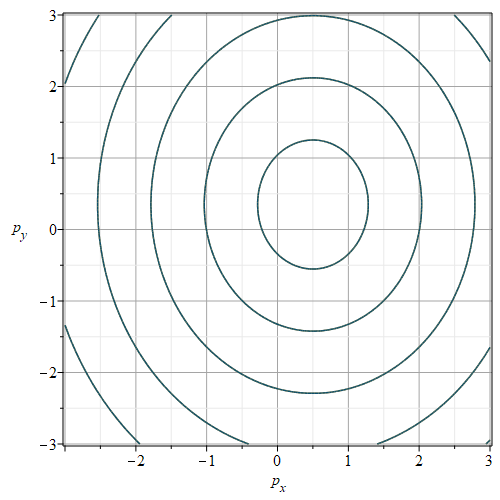}
{\footnotesize\textbf{(h)} Contour plots ($a^{yy}=0.25$, $d_x=0.50$, $d_y=0.35$).}
\end{minipage}
\begin{minipage}{0.30\textwidth}
\centering
\includegraphics[width=\linewidth]{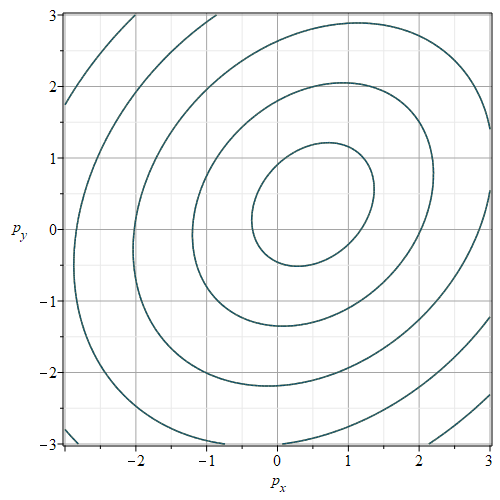}
{\footnotesize\textbf{(i)} Contour plots ($a^{xy}=0.25$, $d_x=0.50$, $d_y=0.35$).}
\end{minipage}

\caption{\small
(a)--(f): Anisotropic deformation of the Dirac spectrum induced by the coefficients $a^{xx}$, $a^{yy}$, and $a^{xy}$, with all other anisotropic parameters set to zero. Unlike $a^{0i}$, these coefficients neither tilt the Dirac cones nor generate a mass gap, but instead modify the effective metric of the dispersion relation. (g)--(i): Anisotropic modifications of the Dirac spectrum induced by the coefficients $a^{xx}$, $a^{yy}$, and $a^{xy}$, with $d_x\neq0$ and $d_y\neq0$. The parameter $d^\mu$ preserves the conical dispersion and acts only by shifting the Dirac point along the momentum direction. Whereas the isotropic case yields circular constant-energy contours, finite values of these coefficients produce elliptic contours with distinct principal axes. Dimensionless units were adopted throughout, with $v_F=1$. Changing the sign of the coefficients rotates the contour pattern by $\pi/2$.
}
\label{fig:all_figures3}

\end{figure}

These results demonstrate that the approximate model provides a reliable framework for investigating the physical role of the anisotropic parameters. Owing to its simpler analytical structure, it allows the influence of each coefficient on the cone geometry, anisotropy, and tilting to be identified more transparently, while retaining the essential phenomenology predicted by the exact formulation. Consequently, the approximate dispersion constitutes a useful tool for exploring the parameter space and establishing qualitative trends before performing a full analysis based on the exact theory. This close qualitative agreement justifies the use of the approximate dispersion throughout this work to interpret the individual effects of the anisotropic parameters, whereas the exact theory serves as the reference for assessing the quantitative accuracy of these predictions.
The spatial anisotropy coefficients $a^{xx}$, $a^{yy}$, and $a^{xy}$ affect the spectrum in a fundamentally different manner, as shown in Fig.~\ref{fig:all_figures3}. Rather than producing a rigid displacement or a cone tilt, these coefficients deform the effective metric of momentum space. As a consequence, the originally circular constant-energy contours become elliptical, with the orientation and eccentricity determined by the particular anisotropic coefficient. The dispersion remains gapless, and no overtilting occurs within the parameter range considered. When combined with finite values of $d_x$ and $d_y$, the anisotropic deformation is preserved while the entire pattern is translated in momentum space, demonstrating that both effects are largely independent and additive.

\begin{figure}[htbp]
\centering

\begin{minipage}{0.32\textwidth}
\centering
\includegraphics[width=\linewidth]{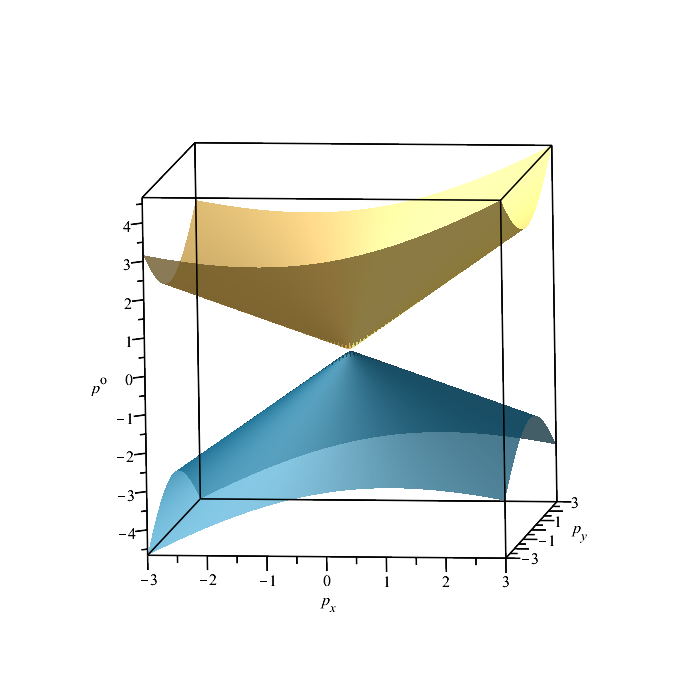}
{\footnotesize\textbf{(a)} Dispersion relation ($a^{xx}=0.30$, $a^{0x}=0.25$).}
\end{minipage}
\begin{minipage}{0.32\textwidth}
\centering
\includegraphics[width=\linewidth]{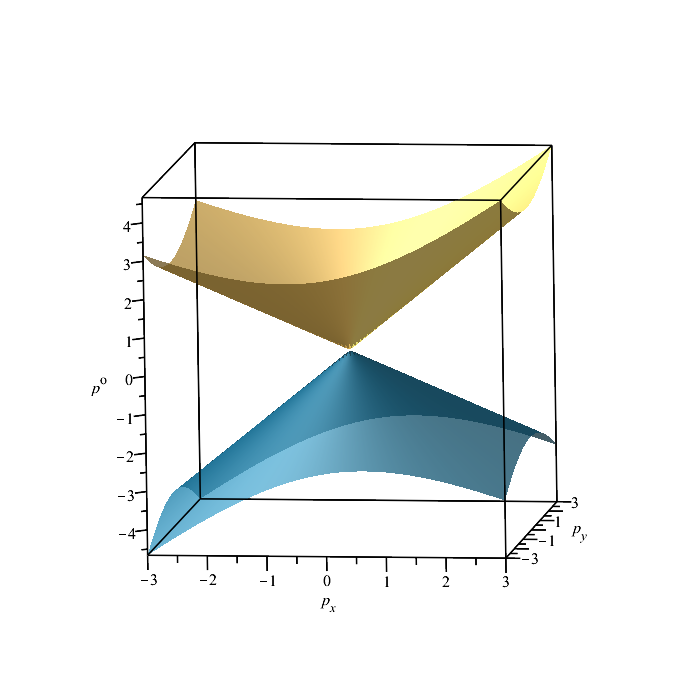}
{\footnotesize\textbf{(b)} Dispersion relation ($a^{yy}=0.30$, $a^{0x}=0.25$).}
\end{minipage}
\begin{minipage}{0.32\textwidth}
\centering
\includegraphics[width=\linewidth]{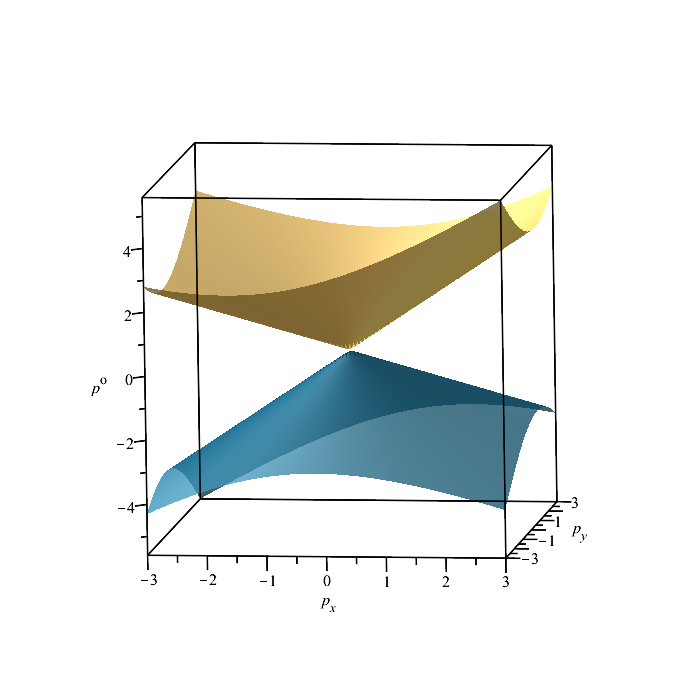}
{\footnotesize\textbf{(c)} Dispersion relation ($a^{xy}=0.30$, $a^{0x}=0.25$).}
\end{minipage}
\vspace{0.1cm}
\begin{minipage}{0.28\textwidth}
\centering
\includegraphics[width=\linewidth]{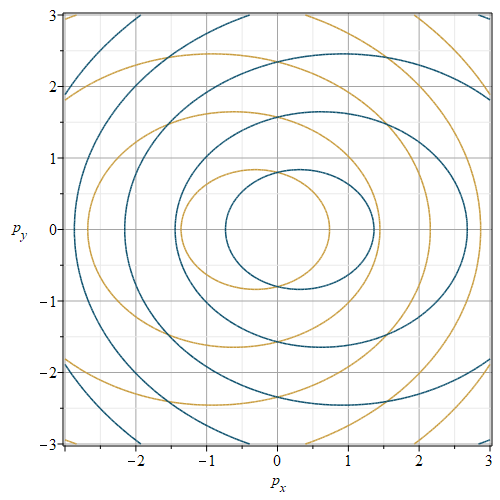}
{\footnotesize\textbf{(d)} Contour plots of (a).}
\end{minipage}
\begin{minipage}{0.28\textwidth}
\centering
\includegraphics[width=\linewidth]{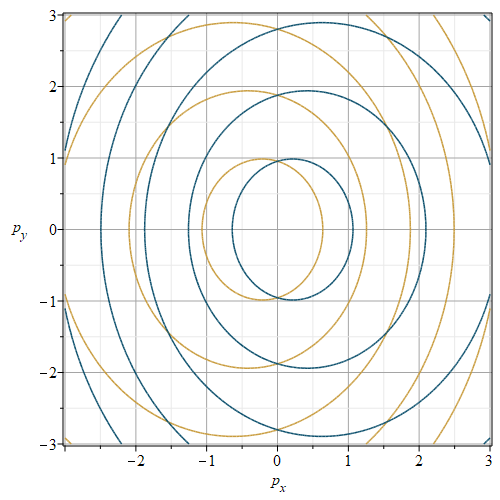}
{\footnotesize\textbf{(e)} Contour plots of (b).}
\end{minipage}
\begin{minipage}{0.28\textwidth}
\centering
\includegraphics[width=\linewidth]{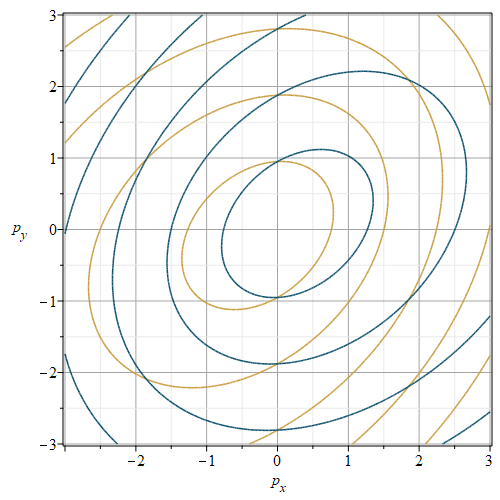}
{\footnotesize\textbf{(f)} Contour plots of (c).}
\end{minipage}
\vspace{0.1cm}
\begin{minipage}{0.28\textwidth}
\centering
\includegraphics[width=\linewidth]{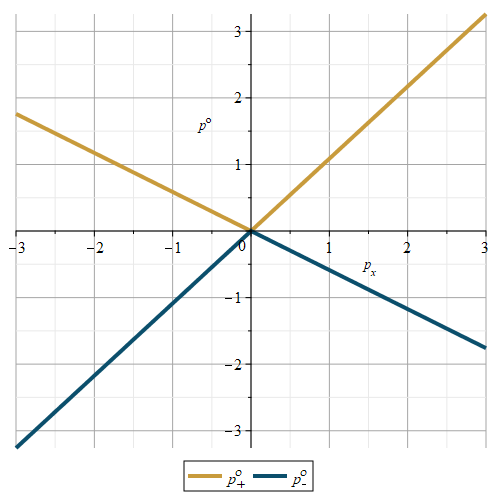}
{\footnotesize\textbf{(g)} Energy cut of (a) along $p_y=0$.}
\end{minipage}
\begin{minipage}{0.28\textwidth}
\centering
\includegraphics[width=\linewidth]{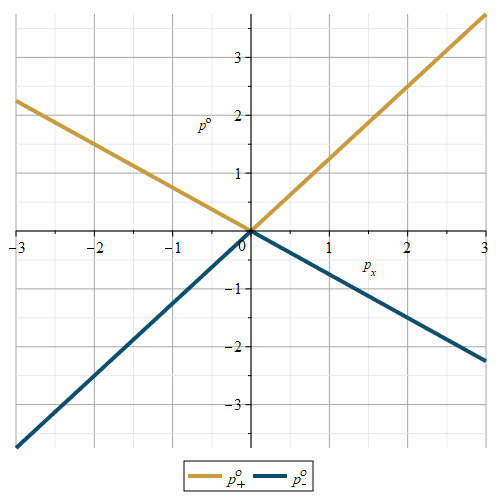}
{\footnotesize\textbf{(h)} Energy cut of (b) along $p_y=0$.}
\end{minipage}
\begin{minipage}{0.28\textwidth}
\centering
\includegraphics[width=\linewidth]{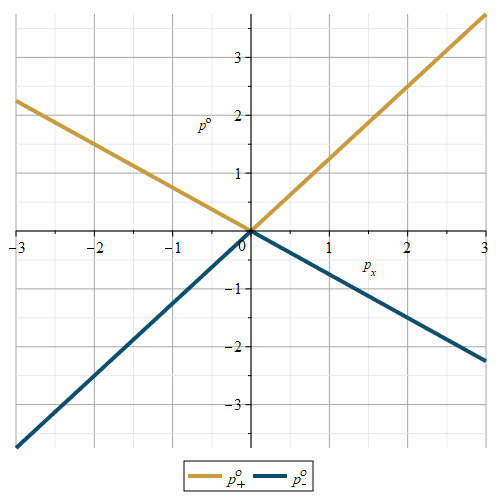}
{\footnotesize\textbf{(i)} Energy cut of (c) along $p_y=0$.}
\end{minipage}
\vspace{0.1cm}
\begin{minipage}{0.28\textwidth}
\centering
\includegraphics[width=\linewidth]{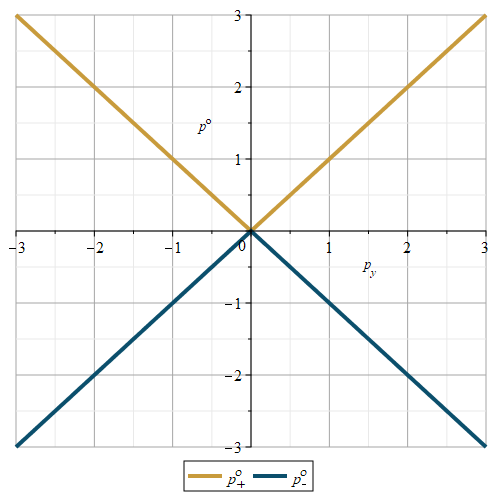}
{\footnotesize\textbf{(j)} Energy cut of (a) along $p_x=0$.}
\end{minipage}
\begin{minipage}{0.28\textwidth}
\centering
\includegraphics[width=\linewidth]{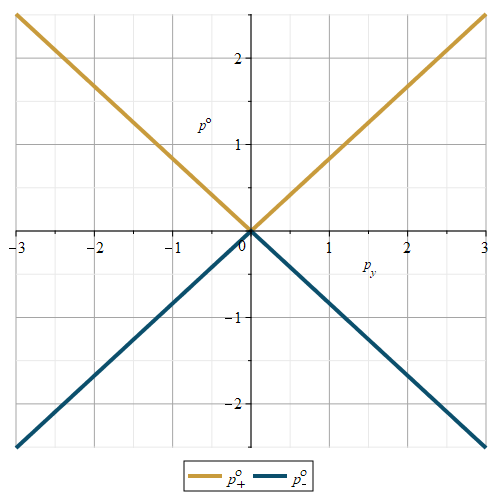}
{\footnotesize\textbf{(k)} Energy cut of (b) along $p_x=0$.}
\end{minipage}
\begin{minipage}{0.28\textwidth}
\centering
\includegraphics[width=\linewidth]{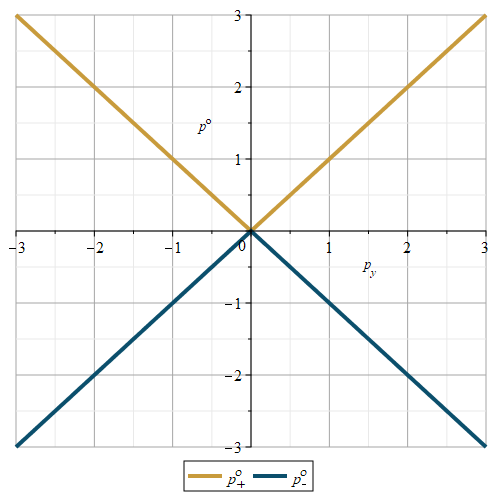}
{\footnotesize\textbf{(l)} Energy cut of (c) along $p_x=0$.}
\end{minipage}

\caption{\small
The dispersion relations of a Type-I Dirac cone deformed by anisotropy are shown in panels (a)--(c). The corresponding constant-energy contours, illustrating the displacement of the conduction- and valence-band contours in opposite directions in reciprocal space, are shown in panels (d)--(f). Panels (g)--(i) display energy cuts along $p_y=0$, while panels (j)--(l) show the corresponding cuts along $p_x=0$.
}
\label{fig:all_figures5}
\end{figure}

\begin{figure}[htbp]
\centering

\begin{minipage}{0.32\textwidth}
\centering
\includegraphics[width=\linewidth]{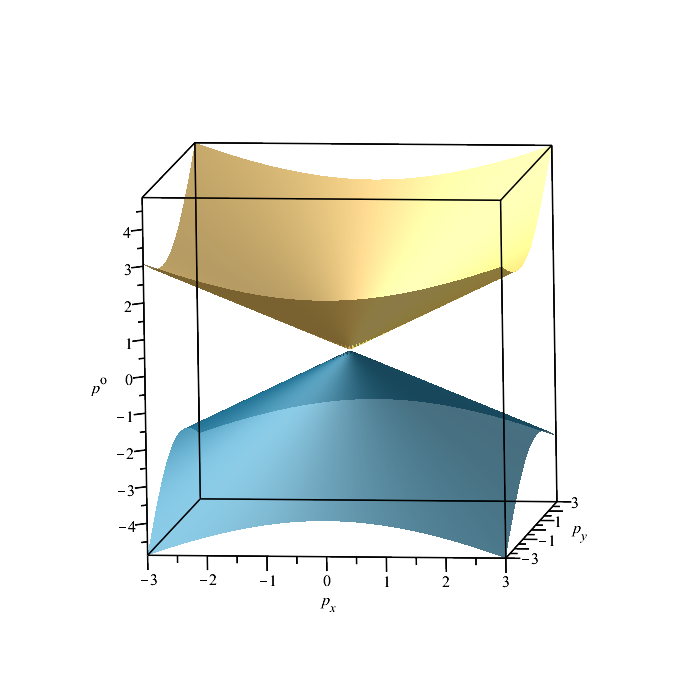}
{\footnotesize\textbf{(a)} Dispersion relation ($a^{xx}=0.25$, $a^{0y}=0.30$).}
\end{minipage}
\begin{minipage}{0.32\textwidth}
\centering
\includegraphics[width=\linewidth]{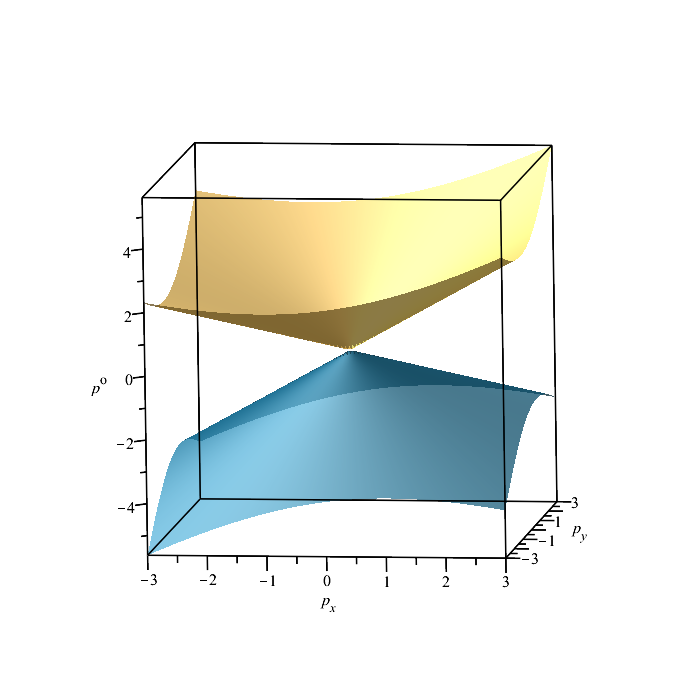}
{\footnotesize\textbf{(b)} Dispersion relation ($a^{xx}=0.25$, $a^{0x}=0.25$, $a^{0y}=0.30$).}
\end{minipage}
\begin{minipage}{0.32\textwidth}
\centering
\includegraphics[width=\linewidth]{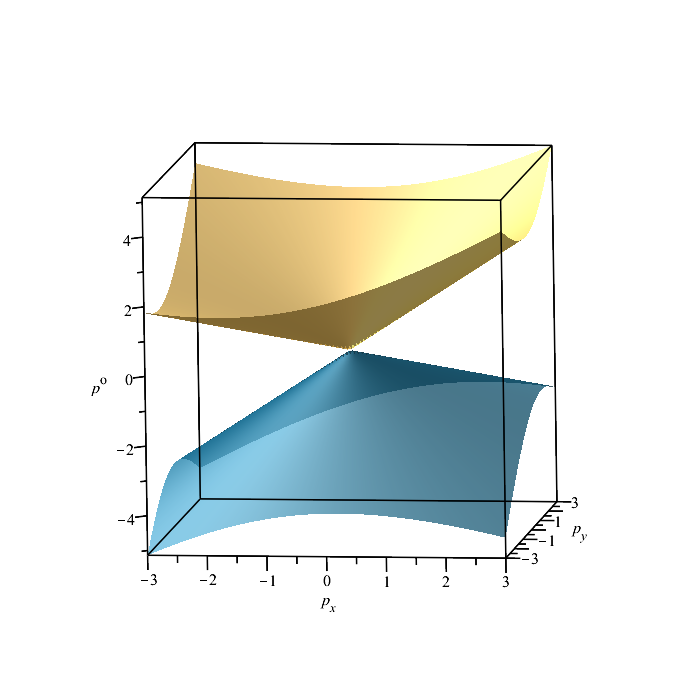}
{\footnotesize\textbf{(c)} Dispersion relation ($a^{xx}=0.25$, $a^{0x}=0.25$, $a^{0y}=0.30$, $a^{xy}=0.20$).}
\end{minipage}
\vspace{0.1cm}
\begin{minipage}{0.28\textwidth}
\centering
\includegraphics[width=\linewidth]{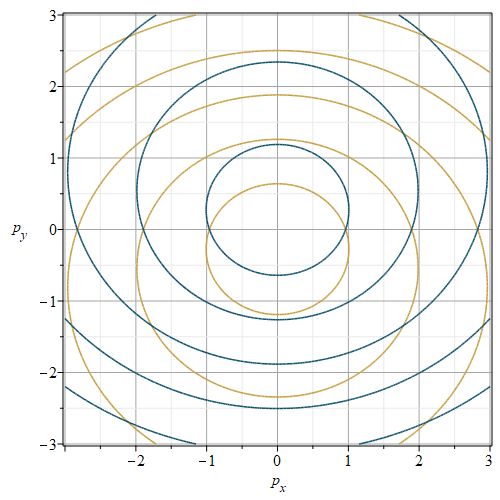}
{\footnotesize\textbf{(d)} Contour plots of (a).}
\end{minipage}
\begin{minipage}{0.28\textwidth}
\centering
\includegraphics[width=\linewidth]{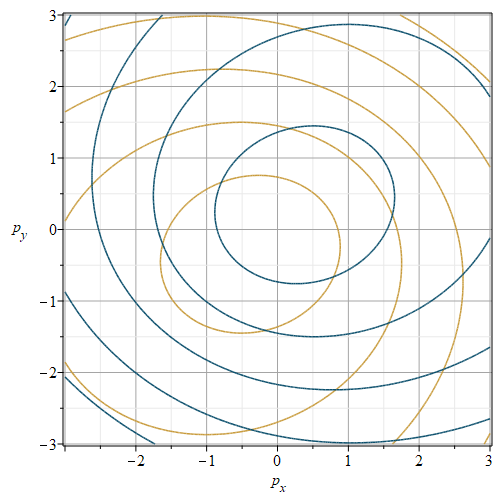}
{\footnotesize\textbf{(e)} Contour plots of (b).}
\end{minipage}
\begin{minipage}{0.28\textwidth}
\centering
\includegraphics[width=\linewidth]{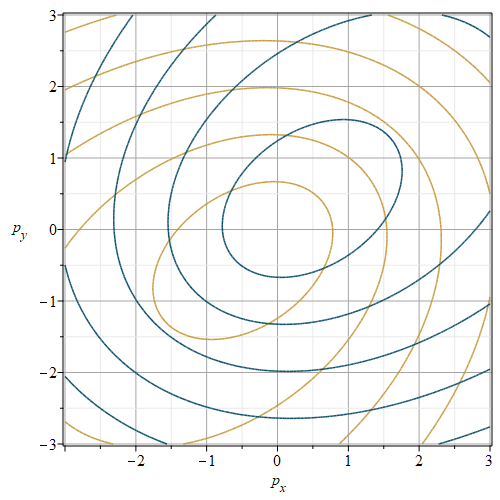}
{\footnotesize\textbf{(f)} Contour plots of (c).}
\end{minipage}
\vspace{0.1cm}
\begin{minipage}{0.28\textwidth}
\centering
\includegraphics[width=\linewidth]{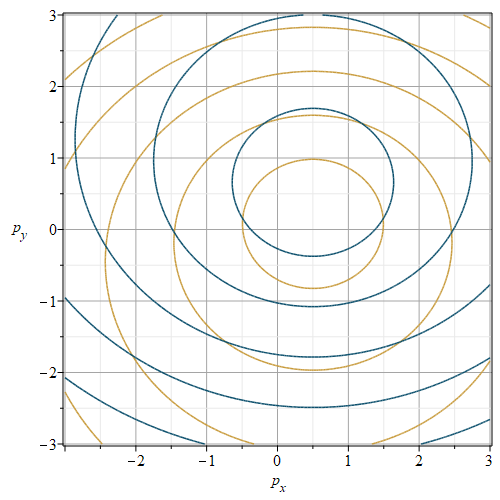}
{\footnotesize\textbf{(g)} Contour plots of (a) ($d_x=0.50$, $d_y=0.35$).}
\end{minipage}
\begin{minipage}{0.28\textwidth}
\centering
\includegraphics[width=\linewidth]{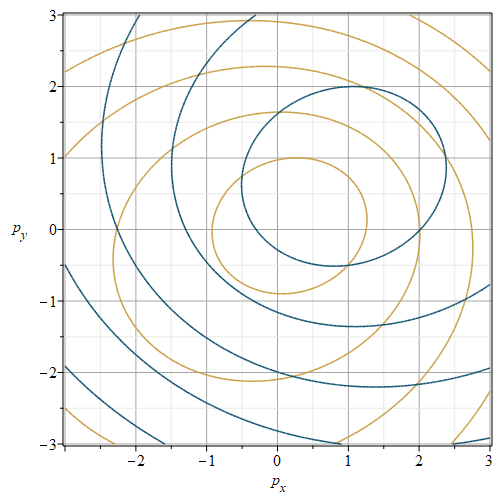}
{\footnotesize\textbf{(h)} Contour plots of (b) ($d_x=0.50$, $d_y=0.35$).}
\end{minipage}
\begin{minipage}{0.28\textwidth}
\centering
\includegraphics[width=\linewidth]{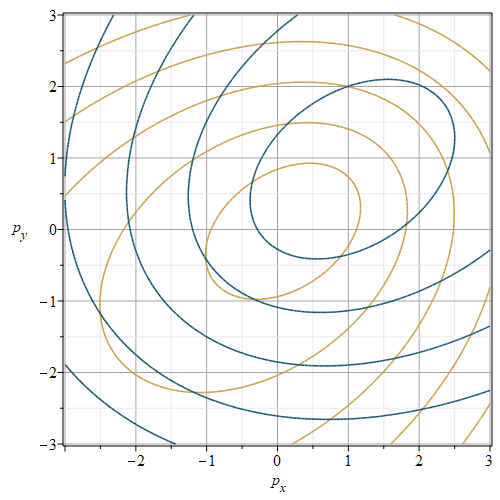}
{\footnotesize\textbf{(i)} Contour plots of (c) ($d_x=0.50$, $d_y=0.35$).}
\end{minipage}
\caption{\small
(a)--(f): Dispersion relations of Dirac cones deformed by tilt and anisotropy together with their corresponding constant-energy contours, showing the displacement of the conduction- and valence-band contours in opposite directions in reciprocal space. Several combinations of the anisotropic parameters $a^{\mu\nu}$ were considered, while all remaining parameters were set to zero. (g)--(i): Anisotropic modifications of the Dirac spectrum with $d_x\neq0$ and $d_y\neq0$. The parameter $d^\mu$ preserves the conical dispersion and acts only by shifting the Dirac point along the momentum direction.
}
\label{fig:all_figures6}
\end{figure}

The interplay between anisotropy and cone tilting is examined in Figs.~\ref{fig:all_figures5} and~\ref{fig:all_figures6}. The simultaneous presence of the coefficients $a^{0i}$ and $a^{ij}$ generates spectra in which the Dirac cones are both tilted and anisotropically deformed. The resulting constant-energy contours exhibit displaced ellipses instead of the circles observed for isotropic systems. Energy cuts taken along the $p_x=0$ and $p_y=0$ directions emphasize the directional dependence introduced by the anisotropy, showing that the deformation modifies the effective velocities differently along distinct momentum directions while the tilt breaks the symmetry between positive and negative energy branches.

The numerical results discussed above can also be understood from an analytical perspective. In particular, the different roles played by the components of the tensor $a^{\mu\nu}$ become evident by examining the dispersion relation in Eq.~\eqref{adim_dispertion}. For instance, by choosing $d^0=d_x=d_y=a^{xy}=m=0$ in Eq.~\eqref{adim_dispertion}, we recover exactly the dispersion relation derived in Ref.~\cite{GOERBIG2} within an effective tight-binding model on an anisotropic triangular lattice with two atoms per unit cell. The determination of the relevant parameters was discussed in Ref.~\cite{MILIĆEVIĆ} using a combination of experimental data from semiconductor microcavities grown by molecular beam epitaxy and a tight-binding model describing the coupling between the \(p_x\) and \(p_y\) orbitals in an isotropic honeycomb lattice. Thus, our theoretical framework not only reproduces known anisotropic effects in two-dimensional Dirac materials, but also unifies them within the structure of the modified Dirac equation in Eq.~(\ref{modEq}). Following the definition of the tilt parameter introduced in Ref.~\cite{GOERBIG2} and using the dispersion relation in Eq.~\eqref{adim_dispertion}, we obtain
\begin{eqnarray}
\tilde{w}_0= \sqrt{\frac{(a^{0x})^2}{1-a^{xx}}+\frac{(a^{0y})^2}{1-a^{yy}}}.    
\end{eqnarray}
In the absence of the components \(a^{0i}\), no tilt is generated. For \(\tilde{w}_0 < 1\), the system exhibits standard or tilted Type-I Dirac cones, whereas \(\tilde{w}_0 > 1\) corresponds to Type-II Dirac cones. The critical case \(\tilde{w}_0 = 1\) describes the transition between Type-I and Type-II regimes and defines the Type-III Dirac cone. The analytical expression for the tilt parameter therefore provides a direct criterion for classifying the different Dirac-cone geometries observed in the numerical results. Combined with the anisotropic deformations governed by the coefficients $a^{ij}$, it offers a unified interpretation of all spectra presented throughout this section.

In summary, these results demonstrate that each class of anisotropic parameters leaves a distinct geometrical fingerprint on the Dirac spectrum. While the coefficients $d^\mu$ act primarily as translation parameters, the coefficients $a^{0i}$ control the cone tilting, and the coefficients $a^{ij}$ govern the anisotropic deformation of the energy landscape. Their combined action provides a versatile framework for engineering the electronic dispersion in two-dimensional Dirac materials. The independent geometric effects associated with each class of parameters are summarized in Table~\ref{tab:geometric_effects}.

\begin{table*}[t]
\caption{\label{tab:geometric_effects}
Independent geometric effects associated with the parameters of the
approximate dispersion relation in Eq.~\eqref{adim_dispertion}.
Combined nonzero parameters generate the corresponding superposition
of these effects.}
\begin{ruledtabular}
\begin{tabular}{cccc}
Parameter
& Contribution to the dispersion
& Main effect
& Geometric signature \\[2pt]
\hline

$d^{0}$
& Additive energy term
& Energy shift
& Vertical translation of the cone \\

$d^{x},\,d^{y}$
& Momentum displacement
& Dirac-point shift
& Translation in momentum space \\

$a^{0x},\,a^{0y}$
& Linear momentum term
& Cone tilt
& Type-I, Type-III, or Type-II cone \\

$a^{xx},\,a^{yy}$
& Diagonal quadratic terms
& Anisotropic deformation
& Elliptic constant-energy contours \\

$a^{xy}$
& Mixed quadratic term
& Off-diagonal anisotropy
& Rotation of the principal axes \\

$m$
& Constant term under the square root
& Gap opening
& Separation of the energy bands \\

Combined parameters
& Simultaneous contributions
& Combined effects
& Shifted, tilted, distorted, gapped cone
\end{tabular}
\end{ruledtabular}
\end{table*}

These geometrical modifications directly affect the electronic~\cite{ALMARZOOG} and optical~\cite{MOJARRORAMIREZ} properties of such materials through changes in the shape, orientation, and topology of the Dirac cones. More importantly, all these physical regimes emerge naturally from the same modified Dirac equation, Eq.~(\ref{modEq}), through different choices of the tensor components $a^{\mu\nu}$. The present formalism therefore provides a unified theoretical framework capable of reproducing and extending the various anisotropic and tilted Dirac spectra reported in the literature.

\FloatBarrier

\section{Validation of the effective model from first-principles calculations}
\label{validation}
After establishing the theoretical framework and analyzing the role of each anisotropic parameter independently, it is important to demonstrate that the proposed model can quantitatively describe realistic materials. In this section, we validate the effective theory by applying it to graphene under uniform biaxial and uniaxial strain using first-principles calculations. Graphene provides an ideal benchmark because its low-energy electronic structure is accurately described by massless Dirac fermions, while its response to both biaxial and uniaxial strain has been extensively investigated experimentally and theoretically
\cite{GUI,farjam2009comment,wong2012strain,ni2008uniaxial,si2016strain}.
Moreover, uniform biaxial strain preserves the lattice symmetry, allowing a direct comparison between the symmetry constraints of the effective model and the electronic structure obtained from Density Functional Theory (DFT) \cite{Martin, KohnSham}. In contrast, uniaxial strain explicitly breaks the in-plane symmetry, providing a complementary test of the model under anisotropic conditions and allowing the strain-induced deformation of the Dirac cones to be quantitatively analyzed.

For each strain value, the electronic band structure around the Dirac point was calculated using DFT, with details of the simulations provided in Appendix~\ref{app:compmet}.
For comparison with the first-principles results, it is convenient to reformulate the dispersion relation derived in Eq.~\eqref{adim_dispertion} in terms of the momentum range used in the fitting procedure. We introduce a cutoff wave vector, $k_0$, and the corresponding energy scale $E_c=\hbar v_F k_0$, where $v_F$ is the graphene Fermi velocity. The momentum coordinates are defined relative to the strain dependent Dirac-point position obtained directly from the DFT band structure,
\begin{equation}
\label{eq:qi}
q_i=\frac{k_i-K_{D,i}}{k_0},
\end{equation}
where $K_{D,i}$ denotes the position of the Dirac point for a given strain. With the dimensionless energy $\tilde{E}=E/E_c$ and $\tilde{E}_0=E_0/E_c$, the dispersion relation in Eq.~\eqref{adim_dispertion} can be written as
\begin{equation}
\tilde{E} = \tilde{E}_0 + a_{0x}q_x + a_{0y}q_y
   \pm \sqrt{(1-a_{xx})\left(q_x - D_x\right)^2 + (1-a_{yy})\left(q_y - D_y\right)^2 + 2a_{xy}\left(q_x - D_x\right)\left(q_y - D_y\right) + \tilde{m}^2},
\label{eq:fit}
\end{equation}
where $\tilde{m}=mv_F/(\hbar k_0)$ and $D_{i} = \frac{v_{F} }{\hbar k_0}d_{i}$. In this formulation, the strain-induced displacement of the Dirac point is separated from the deformation of its local dispersion. The former is determined directly from the DFT band structure through $\Delta\mathbf{K}_D=\mathbf{K}_D(\epsilon)-\mathbf{K}_D(0)$, whereas the parameters $a_{0x}$, $a_{0y}$, $a_{xx}$, $a_{yy}$, $a_{xy}$, and $\tilde{m}$ characterize the local electronic structure around the displaced Dirac point. This formulation is particularly suitable for fitting the low-energy DFT bands, since $k_0$ is directly determined by the fitting region and all remaining parameters are dimensionless. The fitting procedure therefore provides a direct correspondence between the first-principles electronic structure and the phenomenological model, while allowing the strain-induced displacement of the Dirac point to be treated independently from the deformation of the Dirac cone.

\begin{figure}[!htb]
\centering
\includegraphics[width=0.9\textwidth]{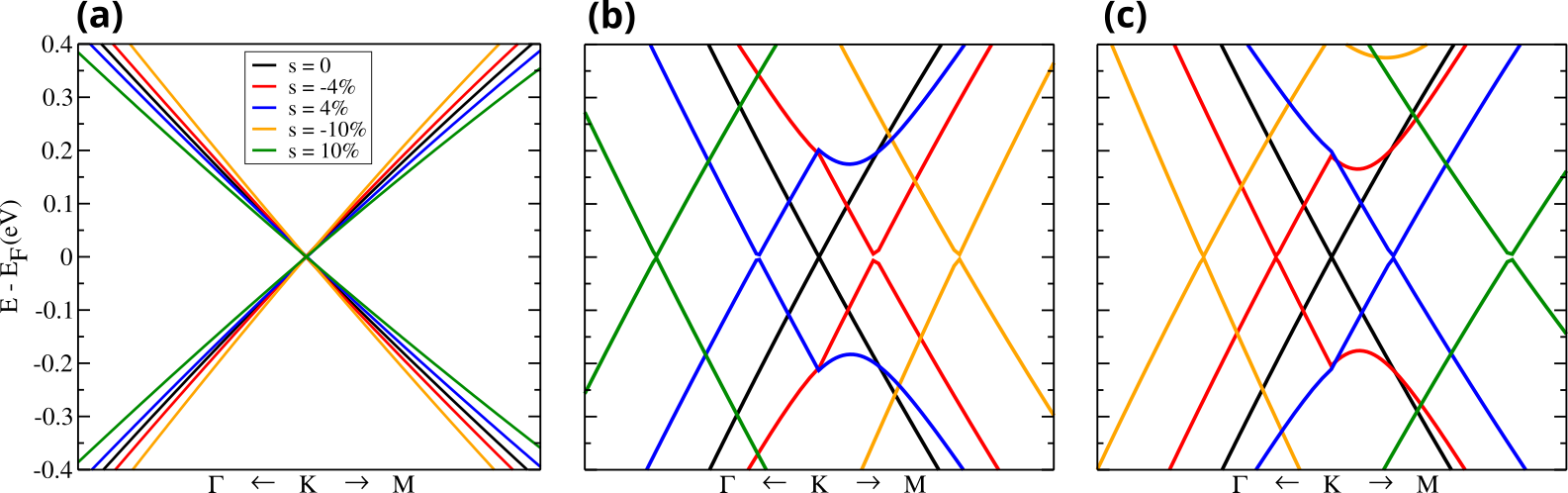}
\caption{Evolution of the electronic band structure around the Dirac point of graphene under different strain conditions. (a) Uniform biaxial strain. (b) Uniaxial strain applied along the zigzag direction. (c) Uniaxial strain applied along the armchair direction.}
\label{fig:bands}
\end{figure}

Fig.~\ref{fig:bands} shows the evolution of the electronic bands around the Dirac point for graphene under different deformation conditions. For biaxial strain [Fig.~\ref{fig:bands}(a)], the Dirac cones remain isotropic, and the main effect of both tensile and compressive strain is a modification of the slope of the valence and conduction bands, reflecting a change in the Fermi velocity while preserving the overall symmetry of the dispersion. In contrast, uniaxial strain breaks the in-plane equivalence of the lattice and produces a markedly anisotropic response. For strain applied along the zigzag direction [Fig.~\ref{fig:bands}(b)], the band dispersion becomes direction dependent, with the slopes and local geometry of the Dirac cones evolving differently under tensile and compressive deformation. A similar anisotropic behavior is observed for uniaxial strain along the armchair direction [Fig.~\ref{fig:bands}(c)], although the modification of the band structure differs from that obtained for zigzag strain due to the distinct orientation of the applied deformation relative to the graphene lattice. In all cases, the Dirac cones remain gapless within the strain range considered, making these systems suitable for testing the ability of the effective model to capture strain-induced modifications of the Dirac spectrum through the extracted anisotropic parameters.

\begin{figure}[!htb]
\centering
\includegraphics[width=0.9\textwidth]{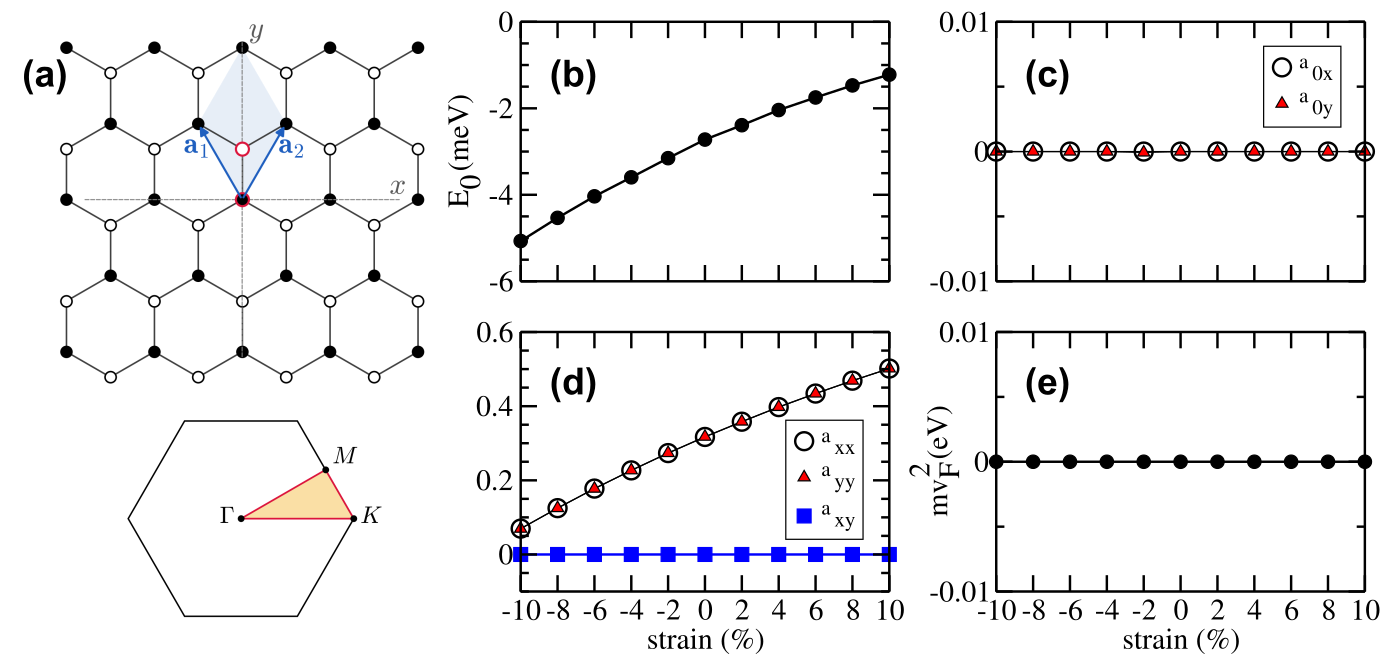}
\caption{(a) Schematic representation of the graphene structure (top) and its first Brillouin zone (bottom). Uniform biaxial strain is applied by simultaneously deforming the lattice along the two in-plane directions, $x$ and $y$, whereas uniaxial strain is applied independently along the $x$ (zigzag) or $y$ (armchair) direction. (b)--(e) Evolution of the fitting parameters extracted from Eq.~(50) as a function of the applied biaxial strain: (b) energy-shift parameter $E_0$; (c) tilt coefficients $a_{0x}$ and $a_{0y}$; (d) anisotropy coefficients $a_{xx}$, $a_{yy}$, and $a_{xy}$; and (e) squared mass parameter $mv_F^{2}$.}
\label{fig:grapBs}
\end{figure}

In Fig.~\ref{fig:grapBs} we present the evolution of the fitting parameters extracted from Eq.~(50) for graphene under uniform biaxial strain. The strain configuration is illustrated schematically in Fig.~\ref{fig:grapBs}(a), where the graphene lattice is simultaneously deformed along the two in-plane directions: zigzag ($x$) and armchair ($y$). Therefore, tensile or compressive strain is applied uniformly to the lattice, preserving its symmetry. 
As shown in Fig.~\ref{fig:grapBs}(b), the energy-shift parameter $E_0$, which appears as an additive constant in Eq.~(50), evolves systematically with the applied strain. Although its variation is approximately linear, the corresponding energy shift remains very small and is therefore nearly negligible compared with the overall energy scale of the Dirac dispersion. This behavior indicates that biaxial deformation produces only a minor rigid displacement of the Dirac spectrum in energy, while its main effect is associated with the modification of the Dirac-cone dispersion.

In contrast, the tilt coefficients $a_{0x}$ and $a_{0y}$ remain essentially zero throughout the investigated strain range, as shown in Fig.~\ref{fig:grapBs}(c). In Eq.~(50), these parameters are associated with the terms that produce an asymmetric linear contribution to the dispersion and, consequently, a tilt of the Dirac cone. Their negligible magnitudes therefore demonstrate that uniform biaxial strain does not induce any measurable cone tilt. This result is fully consistent with the symmetric deformation illustrated in Fig.~\ref{fig:grapBs}(a), which preserves the equivalence between the in-plane directions and does not introduce a preferred direction for the Dirac-cone tilt.
The coefficients $a_{xx}$, $a_{yy}$, and $a_{xy}$ determine the strain-induced modification of the Dirac-cone geometry. As shown in Fig.~\ref{fig:grapBs}(d), the diagonal coefficients $a_{xx}$ and $a_{yy}$ evolve identically with strain, whereas the off-diagonal coefficient $a_{xy}$ remains negligible over the entire strain interval. The relation $a_{xx}\approx a_{yy}$ indicates that biaxial strain modifies the Dirac dispersion isotropically, preserving the equivalence between the $x$ and $y$ directions. At the same time, the negligible value of $a_{xy}$ indicates that the principal axes of the constant-energy contours do not undergo a significant rotation. The extracted parameters therefore reproduce the geometrical modification expected from the symmetry of uniformly strained graphene.

Finally, Fig.~\ref{fig:grapBs}(e) shows that the fitted squared mass parameter, $mv_F^{2}$, remains essentially zero for all applied strains. Since this parameter controls the gap-opening contribution in Eq.~(50), its negligible value confirms that uniform biaxial strain does not induce a gap in graphene within the investigated strain range. Thus, the characteristic gapless semimetallic nature of graphene is preserved under the applied biaxial deformation.

\begin{figure}[!htb]
\centering
\includegraphics[width=1.0\textwidth]{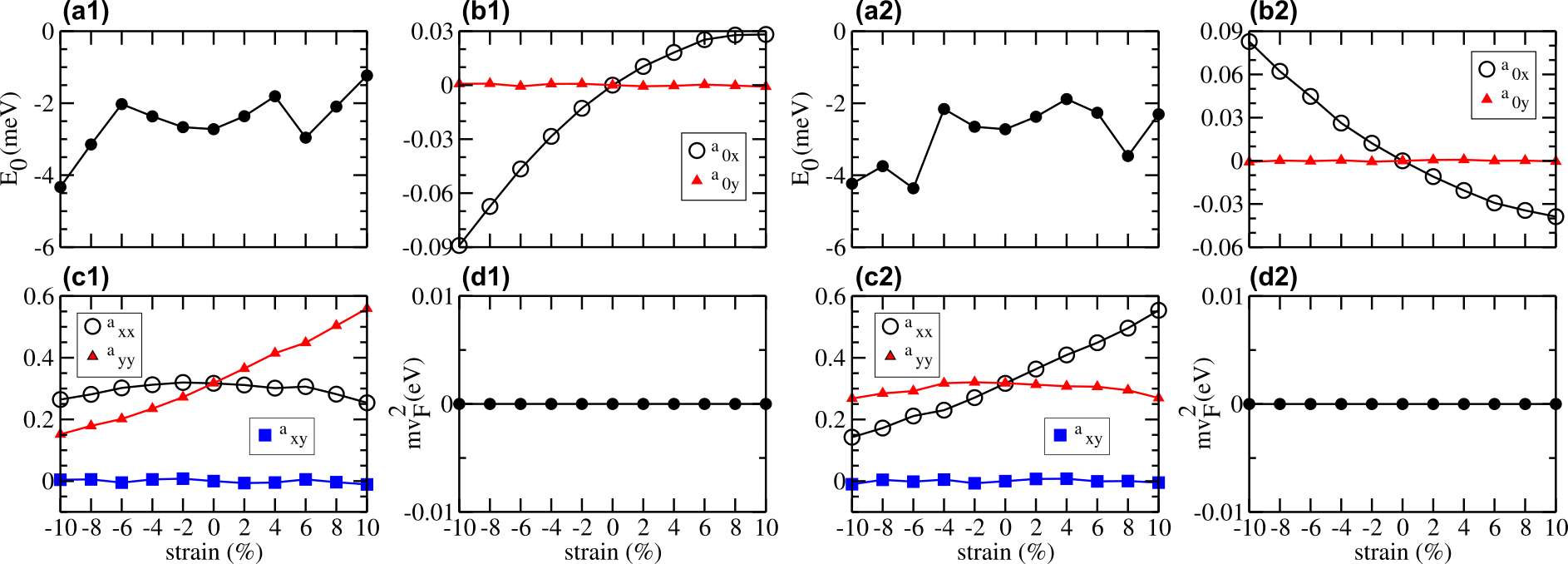}
\caption{Evolution of the fitting parameters extracted from Eq.~(50) for graphene under uniaxial strain applied along the zigzag and armchair directions. Panels (a1)--(d1) correspond to uniaxial strain along the zigzag direction, whereas panels (a2)--(d2) correspond to uniaxial strain along the armchair direction.}
\label{fig:grapUni}
\end{figure}

Fig.~\ref{fig:grapUni} summarizes the fitting parameters obtained from Eq.~(50) for graphene subjected to uniaxial strain along the zigzag and armchair directions. As illustrated schematically in Fig.~\ref{fig:grapBs}(a), the uniaxial deformation is applied along a single crystallographic direction, thereby breaking the equivalence between the two in-plane directions. This reduction of symmetry is reflected directly in the fitted parameters and provides a complementary test of the effective model under an intrinsically anisotropic deformation.

For uniaxial strain along the zigzag direction [Figs.~\ref{fig:grapUni}(a1)--(d1)], the energy-shift parameter $E_0$ exhibits only a relatively small variation, remaining on the meV scale throughout the investigated strain range. Thus, although the absolute energy of the Dirac spectrum is modified by the deformation, the corresponding rigid energy displacement remains small compared with the characteristic energy scale of the bands. More pronounced changes are observed in the tilt sector. The coefficient $a_{0x}$ varies continuously with strain, changing from negative values under compressive strain to positive values under tensile strain, whereas $a_{0y}$ remains essentially zero. This behavior indicates that uniaxial zigzag strain induces a finite, strain-dependent tilt of the Dirac cone, predominantly associated with the $x$ component of the linear momentum term in Eq.~(50). Nevertheless, the magnitude of $a_{0x}$ remains small over the entire strain range, indicating that the resulting tilt is weak and that the system remains well within the type-I Dirac cone regime.

The strongest effect of zigzag strain is observed in the spatial anisotropy coefficients $a_{xx}$ and $a_{yy}$ [Fig.~\ref{fig:grapUni}(c1)]. While $a_{xx}$ exhibits only a moderate variation with strain, $a_{yy}$ changes substantially, increasing from approximately $0.15$ under strong compression to more than $0.5$ under strong tension. The markedly different evolution of these two coefficients demonstrates that the deformation renormalizes the Dirac dispersion differently along the two crystallographic directions. In contrast, the off-diagonal coefficient $a_{xy}$ remains close to zero, indicating that the principal axes of the anisotropic dispersion remain approximately aligned with the crystallographic directions and that no significant rotation of the principal axes is induced. Finally, $mv_F^{2}$ remains essentially zero [Fig.~\ref{fig:grapUni}(d1)], confirming that zigzag strain does not open an appreciable gap and that the gapless character of graphene is preserved.

A similar qualitative picture emerges for uniaxial strain along the armchair direction [Figs.~\ref{fig:grapUni}(a2)--(d2)], but with a distinct redistribution of the anisotropic response between the two spatial directions. The energy-shift parameter $E_0$ remains on the meV scale and shows no dominant systematic trend, indicating that the rigid displacement of the spectrum is again relatively small. The coefficient $a_{0x}$ displays a pronounced strain dependence, evolving from positive values under compression to negative values under tension, while $a_{0y}$ remains essentially zero. Thus, as in the zigzag case, uniaxial strain generates a finite but relatively weak cone tilt whose magnitude and sign depend systematically on the applied strain. The opposite evolution of $a_{0x}$ compared with the zigzag case further highlights the sensitivity of the effective parameters to the orientation of the applied deformation.
The anisotropy coefficients show the clearest distinction between the two uniaxial strain configurations. For armchair strain, $a_{xx}$ exhibits a strong and approximately monotonic increase with tensile strain, whereas $a_{yy}$ remains comparatively close to a nearly constant value around $0.3$. This behavior is complementary to that observed for zigzag strain, where the dominant variation occurs in $a_{yy}$. 
Interestingly, the strongest renormalization of the Dirac dispersion occurs in the momentum direction transverse to the applied uniaxial strain. For zigzag strain, $a_{yy}$, associated with the armchair momentum direction, exhibits the largest variation, whereas for armchair strain the strongest variation occurs in $a_{xx}$, associated with the zigzag momentum direction. This directional dependence reflects the anisotropic modification of the low-energy electronic structure resulting from the strain-induced changes in the C--C bonding environment.
As in the zigzag case, $a_{xy}$ remains close to zero, indicating that the principal axes of the Dirac cone remain approximately aligned with the crystallographic directions. The fitted $mv_F^{2}$ is also essentially zero throughout the strain range, confirming the absence of a significant strain-induced gap.

The comparison between the two uniaxial configurations therefore reveals a clear directional dependence of the effective Dirac spectrum. While both zigzag and armchair strains generate a finite and strain-dependent tilt through $a_{0x}$ and produce anisotropic renormalization of the Dirac dispersion, the dominant spatial coefficient is different for the two deformation directions, with $a_{yy}$ varying most strongly under zigzag strain, whereas $a_{xx}$ exhibits the largest variation under armchair strain. At the same time, the persistent smallness of $a_{xy}$ and $\tilde{m}^{2}$ indicates that the principal axes remain approximately fixed and that no gap opening occurs. These results demonstrate that the parameters extracted from Eq.~(50) retain a direct correspondence with the crystallographic direction of the applied strain and provide a quantitative description of the resulting anisotropic deformation of the Dirac cone.

The results obtained for strained graphene demonstrate that the proposed effective theory provides a practical framework for connecting microscopic electronic-structure calculations with a geometrical description of Dirac cones. Although graphene provides a particularly clean benchmark, the construction is not restricted to its honeycomb lattice. In principle, the same parameter-extraction procedure can be applied to other two-dimensional Dirac materials, including systems with intrinsic anisotropy or anisotropies induced by strain, substrates, or other external perturbations. In this way, first-principles calculations or experimental band structures could be used to determine the effective parameters and subsequently employ the model to investigate and compare the low-energy electronic properties of different Dirac materials within a common framework.

\section{Concluding Remarks}
\label{conclusion}

Anisotropies in two-dimensional Dirac materials play a central role in determining their electronic properties. Their theoretical description is often formulated in terms of microscopic models or first-principles calculations specifically adapted to a given material or lattice configuration. In this work, we have proposed a material-independent effective model that captures, within a unified framework, a broad class of anisotropic effects that may occur in two-dimensional Dirac systems. The central idea is that, even in the presence of anisotropies, the low-energy behavior of quasiparticles around the Dirac points can still be described effectively by a covariant Dirac equation, provided that its structure is appropriately generalized.

The formulation developed here makes explicit how different geometric modifications of the Dirac cones emerge from distinct sectors of the effective theory. The parameters $d^{\mu}$ describe translations of the Dirac cones in momentum and energy space, the components $a^{0i}$ control their tilting, the spatial components $a^{ij}$ determine anisotropic deformations of the constant-energy contours, and the mass parameter $m$ produces a gap between the valence and conduction bands. Their combined action allows shifted, tilted, distorted, and gapped Dirac cones to be described within the same effective framework. In this sense, the model plays a role analogous to that of the massless Dirac equation in pristine graphene: once the effective parameters characterizing the low-energy regime are known, the resulting Dirac dynamics can be investigated without requiring the microscopic structure of the material to be explicitly introduced at every stage.

The comparison with first-principles calculations for strained graphene provides a concrete illustration of this phenomenological interpretation. The parameters extracted from the electronic band structure reproduce the symmetry properties expected for biaxial and uniaxial deformations and allow the different strain-induced modifications of the Dirac spectrum to be identified directly in terms of the effective coefficients. This application demonstrates that the parameters of the model can be quantitatively connected with a microscopic description while retaining a transparent geometrical interpretation. Graphene was used here as a benchmark, but the construction itself does not rely on the microscopic structure of graphene and can, in principle, be applied to other two-dimensional Dirac materials whenever their low-energy spectrum is governed by Dirac-like quasiparticles.

The present analysis has been restricted essentially to the free low-energy theory and to the geometric properties of the corresponding dispersion relation. This limitation also indicates a natural direction for further developments. Since the model is formulated at the Lagrangian level in a covariant form, interactions can be incorporated within the usual methods of quantum field theory. Once the effective parameters are determined from experiments or microscopic calculations, they may provide the low-energy input for investigations of dynamical effects, response functions, and quantum corrections in anisotropic Dirac systems. In particular, a quantum field theoretical extension would make it possible to study how the anisotropic parameters evolve in the presence of interactions and how they contribute to physical observables beyond the single-particle dispersion relation.

In summary, the model presented here provides a general effective description of anisotropic two-dimensional Dirac materials, separating the microscopic origin of the anisotropies from their low-energy geometrical consequences. By connecting a covariant field-theoretical formulation with parameters that can be extracted from realistic electronic structures, the present approach establishes a framework in which different anisotropic phenomena can be described, compared, and eventually extended beyond the free regime within a common theoretical setting.

\appendix

\section{\label{proofminimum}Proof of the minimum of $\mathcal{R}(p_{x}, p_{y})$}

\label{app:minimum}

The general form of the dispersion relation is (see Eq.~\eqref{CompReldispersion})
\begin{equation}
p^{0}=v^{0}+\vec{p}\cdot\vec{w}\pm\sqrt{\mathcal{R}(p_{x},p_{y})}.
\end{equation}
From Eq.~\eqref{dispersionSquared}, we have
\begin{equation}
\mathcal{R}(p_{x},p_{y})=\left({\bf P}-{\bf P}_{0}\right)^{T}\mathcal{A}\left({\bf P}-{\bf P}_{0}\right)-\mathcal{V}^{T}\mathcal{A}^{-1}\mathcal{V}+\mu^{2},
\label{defRAppend}
\end{equation}
where (see Eq.~\eqref{muSquared})
\begin{equation}
\mu^{2}=\frac{\vec{d}^{2}-\left(d^{0}\right)^{2} + m^2}{\left(1+\Upsilon^{00}\right)}v_{F}^{2}+\left(v_{0}\right)^{2}.
\label{musquaredApp}
\end{equation}
The matrix $\mathcal{A}$ is defined by 
\begin{equation}
D \mathcal{A} = S + \frac{1}{D} w w^{T} 
\label{matrixAppendix}
\end{equation}
where, for simplicity, we are using the notation 
\begin{eqnarray}
w=\left(\begin{array}{c}
w_{x}\\
w_{y}
\end{array}\right)&,& w^{T}=\left(\begin{array}{cc}
w_{x} & w_{y}\end{array}\right)\\
S_{ij} &=& \delta_{ij}-\Upsilon^{ij}\\
D &=& 1+ \Upsilon^{00}.
\end{eqnarray}
The components of $w$ are those defined in Eq.~\eqref{wdef}, while $S_{ij}$ is obtained from the definition of $\mathcal{A}_{ij}$ given in Eq.~\eqref{Amatrix}. Thus, in the perturbative approximation we are interested in, the components of $\mathcal{A}$ read
\begin{eqnarray}
\mathcal{A}_{xx}\approx 1-2a^{00}-2a^{xx},\\
\mathcal{A}_{yy}\approx 1-2a^{00}-2a^{yy},\\
\mathcal{A}_{xy}\approx \mathcal{A}_{yx}=-2a_{xy},
\end{eqnarray}
where we have neglected second-order terms in $a^{\mu\nu}$. Since we have $\mathcal{A}_{xx} >0$ and $\bf{det}\mathcal{A} > 0$ in this regime, the matrix $\mathcal{A}$ is positive definite. Therefore,  the quadratic form $\left({\bf P}-{\bf P}_{0}\right)^{T}\mathcal{A}\left({\bf P}-{\bf P}_{0}\right)$ is non-negative, and $\mathcal{R}(p_{x},p_{y})$ has a global minimum located at ${\bf P}_{0}$. Let us show that this minimum is \begin{equation}
 \mathcal{R}_{min} = v_{F}^{2}\frac{m^2}{D}. 
 \label{minimimAppend}
\end{equation} 

From the expression in Eq.~\eqref{defRAppend}, the minimum $\mathcal{R}_{min}$ is given by
\begin{equation}
\mathcal{R}_{min}= -\,\mathcal{V}^{T}\mathcal{A}^{-1}\mathcal{V}+\mu^{2}.   \label{VTAVAppend} 
\end{equation}
Thus, we must show that 
\begin{equation}
\mathcal{V}^{T}\mathcal{A}^{-1}\mathcal{V} = \frac{\vec{d}^{2}-\left(d^{0}\right)^{2}}{\left(1+\Upsilon^{00}\right)}v_{F}^{2}+\left(v_{0}\right)^{2}.
\label{matrixRelation}
\end{equation}
As we will see, this is a  general result which is independent of the perturbative approximation and emerges from the covariant structure of the model.

First, we note that  the matrix in Eq.~\eqref{matrixAppendix} is the Schur complement \cite{ZHANG} of the $3 \times 3$ matrix
\begin{equation}
G= L \eta L^{T} = \left(\begin{array}{cc}
D & w^{T}\\
w & -S
\end{array}\right),
\label{DefmatrixG}
\end{equation}
Furthermore,  $\eta$ is the matrix representation of the Minkowski metric in (2+1) dimensions and the matrix $L$ is defined from   the three-vector $\Theta^{\mu} = L_{\,\,\sigma}^{\mu}d^{\sigma} = \left( \delta_{\,\,\sigma}^{\mu}+a^{\mu\rho}\eta_{\rho\sigma}\right) d^{\sigma}$ (see Eq.~\eqref{theta} for $b^{\mu} = 0$). Namely,
\begin{equation}
\Theta=\left(\begin{array}{c}
\Theta^{0}\\
\Theta^{x}\\
\Theta^{y}
\end{array}\right) = L d,
\label{matrixTheta}
\end{equation}
with
\begin{equation}
L=\left(\begin{array}{ccc}
1+a^{00} & -a^{0x} & -a^{0y}\\
a^{0x} & 1-a^{xx} & -a^{xy}\\
a^{0y} & -a^{xy} & 1-a^{yy}
\end{array}\right)\qquad\text{and}\qquad d=\left(\begin{array}{c}
d^{0}\\
d_{x}\\
d_{y}
\end{array}\right).
\end{equation}
Here we used that the components $a^{\mu\nu}$ are symmetric.

The inverse of the matrix $G$ in Eq.~\eqref{DefmatrixG} is given by \cite{ZHANG} 
\begin{equation}
G^{-1}=\left(\begin{array}{cc}
D^{-1}-D^{-3}w^{T}\mathcal{A}^{-1}w & +D^{-2}w^{T}\mathcal{A}^{-1}\\
D^{-2}\mathcal{A}^{-1}w & -D^{-1}\mathcal{A}^{-1}
\end{array}\right),
\label{inverseG}
\end{equation}
where  $\mathcal{A}^{-1}$ is the inverse of the matrix $\mathcal{A}$.

Now, using this representation of $\Theta$ in terms of the components $\Theta^{\mu}$, given in Eq.~\eqref{matrixTheta}, and $G^{-1}$ given in Eq.~\eqref{inverseG}, we can show that 
\begin{equation}
\Theta^{T}G^{-1}\Theta=\frac{1}{D}\left(\Theta^{0}\right)^{2}-\frac{1}{D}\left(\Theta^{i}-\frac{\Theta^{0}w_{i}}{D}\right)\left(\mathcal{A}^{-1}\right)_{ij}\left(\Theta^{j}-\frac{\Theta^{0}w_{j}}{D}\right).
\end{equation}
Multiplying this last equation by $\left(\frac{v_{F}}{D}\right)^{2}$ we obtain
\begin{equation}
\frac{v_{F}^{2}}{D}\Theta^{T}G^{-1}\Theta=\left(v_{0}\right)^{2}-v_{i}\left(\mathcal{A}^{-1}\right)_{ij}v_{j},\label{finalThetaG}
\end{equation}
where we have used the definitions of $v_{0}$ and $v_i$ given in Eqs.~\eqref{definitionsParameters0} and \eqref{Vmatrix}, respectively. Note that the second term on  the right-hand side of Eq.~\eqref{finalThetaG} is exactly $\mathcal{V}^{T}\mathcal{A}^{-1} \mathcal{V} $ that we need to calculate  (see Eq.~\eqref{matrixRelation}).

From Eq.~\eqref{DefmatrixG} we see that 
\begin{equation}
L^{T}G^{-1}L=\eta,
\end{equation}
since $\eta^{-1}=\eta$. Thus
\begin{equation}
\Theta^{T}G^{-1}\Theta=d^{T}\left(L^{T}G^{-1}L\right)d = d^{T}\eta d = \left(d^{0}\right)^2 - \vec{d}^2.
\end{equation}
Comparing this result with Eq.~\eqref{finalThetaG}, we obtain Eq.~\eqref{matrixRelation}. Finally, using Eqs.~\eqref{musquaredApp} and \eqref{matrixRelation} in Eq.~\eqref{VTAVAppend}, we obtain Eq.~\eqref{minimimAppend}.

\section{\label{app:compmet} Computational methods}

First-principles calculations were performed using Density Functional Theory (DFT) as implemented in the Quantum ESPRESSO package \cite{GIANNOZZI}. 
A plane-wave basis set was employed in conjunction with the projector-augmented wave (PAW) method \cite{KRESSE}.
The exchange-correlation functional was treated within the generalized
gradient approximation (GGA), using the Perdew–Burke–Ernzerhof
(PBE) formulation \cite{PERDEW}. 
The Kohn–Sham orbitals were expanded
using a plane-wave energy cutoff of 48 Ry. Atomic geometries were
relaxed until the forces on each atom were below 1~meV/\AA. Brillouin
zone sampling was carried out using a $20\times 20\times 1$  Monkhorst–Pack
k-points mesh \cite{MONKHORST}.

To quantify the strain-induced modifications of the Dirac cones, the
DFT band structures were fitted to the dimensionless dispersion relation
given in Eq.~\eqref{eq:fit}. For each strain value, the valence- and conduction-band
energies were sampled on the two-dimensional momentum grid generated for
the band interpolation. The position of the Dirac point,
$\mathbf{K}_{D}=(K_{D,x},K_{D,y})$, was first determined directly from the
DFT bands as the momentum point at which the energy separation between the
conduction and valence bands is minimized. The momentum coordinates were
then expressed relative to this strain-dependent Dirac point according to
Eq.~\eqref{eq:qi}. This procedure allows the displacement of the Dirac point in
momentum space to be separated from the modification of the local
dispersion, as discussed in Sec.~\ref{validation}.

A circular fitting region of radius
$k_0=0.10~\mathrm{\AA}^{-1}$ centered at $\mathbf{K}_{D}$ was used for all
strain values. Keeping the same value of $k_0$ throughout the calculations
ensures that the fitted parameters are obtained from equivalent
momentum-space regions and that their strain dependence is not affected by
changes in the fitting window. 
The nine parameters ($\widetilde{E}_0,a_{0x},a_{0y},a_{xx},a_{yy},a_{xy},
D_x,D_y,\widetilde{m}$)
were determined simultaneously by nonlinear least-squares fitting of the
conduction and valence bands to Eq.~\eqref{eq:fit}. Specifically, the residual
function was constructed by concatenating the differences between the
calculated and fitted valence-band energies with the corresponding
differences for the conduction band. Thus, both branches of the Dirac
dispersion were fitted simultaneously using the same set of parameters.
The mass parameter was constrained to $\widetilde{m}\geq0$, since it enters
the dispersion only through $\widetilde{m}^{\,2}$. The remaining parameters
were left unconstrained during the optimization.


\end{document}